\documentclass[a4paper,11pt]{article}
\pdfoutput=1 % if your are submitting a pdflatex (i.e. if you have
\usepackage{jheppub_mod} % for details on the use of the package, please
\usepackage[T1]{fontenc} % if needed

\usepackage{amsthm} %for theorems and demonstrations 

\renewcommand\Re{\operatorname{Re}}
\renewcommand\Im{\operatorname{Im}}
\usepackage{amsmath,amsthm,verbatim,amssymb,amsfonts,amscd,graphicx,collectbox,mathtools,float}

\usepackage{tikz}
\usetikzlibrary{math}
\usetikzlibrary{arrows,shapes,positioning}
\usetikzlibrary{decorations.markings}
\tikzstyle arrowstyle=[scale=1]
\tikzstyle directed=[postaction={decorate,decoration={markings,
    mark=at position .65 with {\arrow[arrowstyle]{stealth}}}}]
\tikzstyle reverse directed=[postaction={decorate,decoration={markings,
    mark=at position .65 with {\arrowreversed[arrowstyle]{stealth};}}}]

\newlength{\mywidth}
\usepackage[colorlinks=true
,urlcolor=blue
,anchorcolor=blue
,citecolor=blue
,filecolor=blue
,linkcolor=blue
,menucolor=blue
,pagecolor=blue
,linktocpage=true
,pdfproducer=medialab
,pdfa=true
]{hyperref}

\title{\boldmath From tree- to loop-simplicity in affine Toda theories I: Landau singularities and their subleading coefficients}

\author[1]{Patrick Dorey,}
\author[1]{Davide Polvara}

\affiliation[1]{Department of Mathematical Sciences, Durham University, Durham DH1 3LE, United Kingdom }

\emailAdd{p.e.dorey@durham.ac.uk}
\emailAdd{davide.polvara@durham.ac.uk}

\abstract{Various features of the even order poles appearing in the S-matrices of simply-laced affine Toda field theories are analysed in some detail. In particular, the coefficients of first- and second-order 
singularities appearing in the Laurent expansion of the S-matrix around a general $2N^{\rm th}$ order pole are derived in a universal way using perturbation theory at one loop. We show how to cut loop diagrams contributing to the pole into particular products of tree-level graphs that depend on the on-shell geometry of the loop; in this way, we recover the coefficients of the Laurent expansion around the pole exploiting tree-level integrability properties of the theory.   The  analysis is independent of the particular simply-laced theory considered, and all the results agree with those obtained in the conjectured  bootstrapped S-matrices of the ADE series of theories.}

\begin{document} 
\maketitle
\flushbottom

%\section{Introduction}
%\label{sec:intro}

%\include{Temporary_Section}

\section{Introduction}
\label{sect_Intro}

Our understanding of quantum field theories (QFTs) has been considerably advanced by the study of integrable models. In a pioneering paper~\cite{Zamolodchikov:1978xm} Zamolodchikov and Zamolodchikov
explained how to conjecture exact expressions for the S-matrices of many massive integrable quantum field theories through an axiomatic approach: the so-called (exact) S-matrix bootstrap. However, a direct connection with standard perturbation theory is in most cases ill-understood.
In the review paper~\cite{Dorey:1996gd}, the fact that integrability should manifest itself in a priori surprising cancellations between Feynman diagrams contributing to production processes in perturbation theory was emphasised, and a systematic approach to the problem at the tree level was undertaken in~\cite{Gabai:2018tmm,Bercini:2018ysh,Patrick_Davide_paper}, where constraints on masses and couplings necessary for the absence of production were given. Though the problem of classifying all the possible bosonic quantum field theories satisfying these constraints remains open, a universal proof of the absence of production for the entire class of affine Toda field theories was found in~\cite{Patrick_Davide_paper}, thereby providing a tree-level proof of their perturbative
integrability. In the following, we move to loop level, and
consider a subclass of these models, the simply-laced affine Toda theories.
The S-matrices of these models have been bootstrapped \cite{Braden:1989bu,a1,a2,a3,a4,a5,a6,a7,Dorey:1990xa, Dorey:1991zp,Fring:1991gh} and have a beautiful universal structure in terms of the
roots and weights of their underlying Lie algebras \cite{Dorey:1990xa, Dorey:1991zp,Fring:1991gh} \footnote{Results for the S-matrices of more subtle cases such as non-simply laced theories \cite{a9,Corrigan:1993xh,Oota:1997un} and supersymmetric models \cite{a10,a11} have also been found, though a geometrical interpretation of their S-matrices in terms of Dynkin diagrams is less straightforward.}. 
At  tree level, this geometrical structure is the reason for the absence of production \cite{Patrick_Davide_paper}, and all the integrability requirements emerge from the geometry of the underlying root system. Even though the bootstrap program has been able by itself to generate expressions for the S-matrices that have passed many different perturbative checks~\cite{Braden:1989bu,Braden:1990wx,Braden:1990qa,Braden:1991vz,Braden:1992gh}, the underlying mechanism at all loops remains unclear.
Since in the ADE series of affine Toda models the absence of production at the tree level completely constrains the masses and the Lagrangian couplings,
these theories look to be good first candidates to address the general problem of perturbative quantum integrability.

In this paper, we consider the specific problem of finding the coefficients of the S-matrix expansion of a generic simply-laced affine Toda theory 
around an arbitrary pole of order $2N$. If we consider just the singular part of the S-matrix we can write a double expansion
\begin{equation}
\label{Expansion_of_S_around_an_arbitrary_2N_order_pole}
S^{\text{(sing)}}_{ab}(\theta)=\sum_{p=1}^{2N} \frac{1}{(\theta-i\theta_0)^p} \Bigl(\frac{\beta^2}{2h}\Bigr)^p \bigl(a_p + b_p \beta^2 + O(\beta^4) \bigr)
\end{equation}
where, for each term of order $p$ in the Laurent expansion around the pole in $\theta$, $a_p$ represents the leading,  and $b_p$ the subleading, coefficient in the coupling expansion. 
In~\eqref{Expansion_of_S_around_an_arbitrary_2N_order_pole}, $\theta$ is the difference between the rapidities of the two particles of types $a$ and $b$ involved in the scattering, $\beta$ is the coupling of the theory and $\theta_0$ the location of the pole under consideration.
In simply-laced affine Toda models the mass ratios do not renormalize at one loop~\cite{Braden:1989bu}; we will assume that this holds to all orders, so that 
the fusing angles do not depend on $\beta$ and therefore the pole position $\theta_0$ is independent of the coupling. The constant $h$, included for convenience, is the Coxeter number of the Lie algebra associated with the model under consideration. The expansion~\eqref{Expansion_of_S_around_an_arbitrary_2N_order_pole} can be easily derived from the conjectured exact S-matrices of the simply-laced affine Toda theories, written as products of certain building blocks \cite{Braden:1989bu}. 
First, we need to expand the bootstrapped S-matrix in small $\beta$; this is indeed the regime in which Feynman diagrams make sense and the comparison is reasonable. Second, we take the limit to the pole position. 
More challenging is to reproduce the surprisingly-simple values for the coefficients yielded by this procedure from perturbation theory: for example,
the direct computation of the leading coefficients $a_P$ and $b_P$ 
for poles of high orders $P=2N$ would require the summation of huge numbers of multiple-loop Feynman diagrams~\cite{Braden:1990wx}. 
Nevertheless, the expansion \eqref{Expansion_of_S_around_an_arbitrary_2N_order_pole} also contains
lower-order terms in $\beta$, which should be reproduced from diagrams with smaller numbers of loops\footnote{In a two to two scattering process, diagrams with $L$ loops contribute to  order $\beta^{2(1+L)}$. This  will become clear in the next section when we will introduce the Lagrangian of the model.} and therefore can be more easily computed. Information on higher-order singularities is therefore contained also in reasonably simple one-loop computations, and these will be our focus below.

As just mentioned, the double expansion of the bootstrapped S-matrices at pole positions 
yields simple and indeed universal results. For 
example, as shown in appendix~\ref{Appendix_S_matrix_expansion}, the coefficients $a_1$ and $b_1$ are always zero no matter the order of the even order pole, while $a_2=N$. 
In this paper, we show how to derive these coefficients for a generic simply-laced affine Toda model using perturbation theory. The result for $a_1$ is easily understood. It corresponds to an order $\beta^2$ contribution to a simple pole in the perturbative expansion, which for a two to two process can only come from on-shell tree-level diagrams with two three-point couplings. However it is easily seen from the fusing rule of \cite{Dorey:1990xa,Dorey:1991zp} that such diagrams are never on shell, neither in the direct nor in the crossed channel, at 
an even order pole position $\theta=i\theta_0$.
More difficult to derive are the coefficients $a_2$ and $b_1$ from Feynman diagram computations,  corresponding to one loop results in perturbation theory.
The coefficient $a_2$ for $N=1$ was computed for the first time in~\cite{Braden:1990wx}, in all the ADE series of models, while $b_1$ was determined, again for $N=1$, in~\cite{Braden:1992gh} though only for the $A_r^{(1)}$ models. Making use of tree-level properties common to all the ADE series of models  we extend these results to any simply-laced affine Toda theory and arbitrary $N$. To determine the values of $a_2$ and $b_1$ we study Landau singularities in one-loop Feynman diagrams since, as pointed out many years ago by Coleman and Thun \cite{Coleman:1978kk}, higher-order poles in the S-matrices of (1+1)-dimensional theories are due to the presence, for particular values of the rapidity, of multiple simultaneously on-shell propagators inside loop diagrams (in higher dimensions Landau singularities lead to branch points, but in two dimensions they give poles). We show how to determine the coefficients of the Laurent expansion at the pole by properly cutting the loops and transforming the loop integrals into products of particular tree-level diagrams presenting internal on-shell bound state propagators. In this manner, we recover the coefficients $a_2$ and $b_1$ using tree-level integrability properties of the theory. The choice of  propagators  to be cut inside the loop is determined by the on-shell geometry of the diagrams and is a key point to 
evaluating the coefficients of the Laurent expansion. 

The rest of this paper is organised as follows. In section \ref{section_tree_level_and_bootstrap} we review the mechanism responsible for the cancellation of $4$-point tree-level non-elastic processes in perturbation theory with a particular focus on the cancellation of poles in 
Feynman diagrams connected by flips of type II, according to the convention used in \cite{Patrick_Davide_paper,Braden:1990wx}. These cancellations are particularly useful to understand the simplification mechanism that manifests itself
at one loop.
In section \ref{Singularities_from_Feynman_diagrams_main_section} we compute, using perturbation theory, the coefficients $a_2$ and $b_1$ in equation~\eqref{Expansion_of_S_around_an_arbitrary_2N_order_pole} showing the emergence of a universal behaviour not depending on the simply-laced theory studied. 
Section~\ref{The_Conclusions} gives our conclusions, discussing the results
obtained and presenting possible generalizations and open problems. 
Appendix~\ref{Appendix_S_matrix_expansion} reviews the building block structure of the bootstrapped S-matrix; in particular, we extract from the bootstrapped result the coefficients $a_2$ and $b_1$ of the Laurent expansion~\eqref{Expansion_of_S_around_an_arbitrary_2N_order_pole} showing a perfect match with the quantities obtained from perturbation theory. Finally, in appendix~\ref{Appendix_contour_integrals}, we justify why the spatial components of the loop momenta have to be purely imaginary when we integrate in the neighbourhood of the Landau singularities.

\section{Pole cancellation at tree level}
\label{section_tree_level_and_bootstrap}
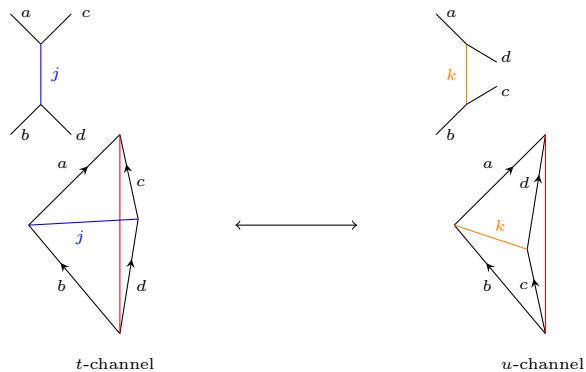
\begin{figure}
\medskip

\begin{center}
\begin{tikzpicture}
\tikzmath{\y=0.8;}

%1st diagram 
\draw[directed] (5.8*\y,0*\y+5*\y) -- (7.3*\y,1.5*\y+5*\y);
\draw[directed] (7.3*\y,-1.8*\y+5*\y) -- (5.8*\y,0*\y+5*\y);
\draw[directed] (7.3*\y,-1.8*\y+5*\y) -- (7.6*\y,0.1*\y+5*\y);
\draw[directed] (7.6*\y,0.1*\y+5*\y) -- (7.3*\y,1.5*\y+5*\y);
\draw[red] (7.3*\y,-1.8*\y+5*\y) -- (7.3*\y,1.5*\y+5*\y);
\draw[blue] (5.8*\y,0*\y+5*\y) -- (7.6*\y,0.1*\y+5*\y);

\filldraw[black] (6.1*\y,1*\y+5*\y)  node[anchor=west] {\tiny{$a$}};
\filldraw[black] (6.1*\y,-1*\y+5*\y)  node[anchor=west] {\tiny{$b$}};
\filldraw[black] (7.4*\y,0.7*\y+5*\y)  node[anchor=west] {\tiny{$c$}};
\filldraw[black] (7.4*\y,-1*\y+5*\y)  node[anchor=west] {\tiny{$d$}};
\filldraw[blue] (6.4*\y,4.8*\y)  node[anchor=west] {\tiny{$j$}};

\filldraw[black] (6.4*\y,0.7*\y+2*\y)  node[anchor=west] {\tiny{$t$-channel}};

\draw[] (5.5*\y,8.5*\y) -- (6*\y,8*\y);
\draw[] (5.5*\y,6.5*\y) -- (6*\y,7*\y);
\draw[blue] (6*\y,8*\y) -- (6*\y,7*\y);
\draw[] (6*\y,8*\y) -- (6.5*\y,8.5*\y);
\draw[] (6.5*\y,6.5*\y) -- (6*\y,7*\y);

\filldraw[black] (5.5*\y,8.5*\y)  node[anchor=west] {\tiny{$a$}};
\filldraw[black] (5.5*\y,6.5*\y)  node[anchor=west] {\tiny{$b$}};
\filldraw[black] (6.5*\y,8.5*\y)  node[anchor=west] {\tiny{$c$}};
\filldraw[black] (6.4*\y,6.5*\y)  node[anchor=west] {\tiny{$d$}};
\filldraw[blue] (6*\y,7.5*\y)  node[anchor=west] {\tiny{$j$}};

%2nd diagram 
\draw[directed] (5.8*\y+7*\y,0*\y+5*\y) -- (7.3*\y+7*\y,1.5*\y+5*\y);
\draw[directed] (7.3*\y+7*\y,-1.8*\y+5*\y) -- (5.8*\y+7*\y,0*\y+5*\y);
\draw[directed] (7*\y+7*\y,-0.4*\y+5*\y) -- (7.3*\y+7*\y,1.5*\y+5*\y);
\draw[directed] (7.3*\y+7*\y,-1.8*\y+5*\y) -- (7*\y+7*\y,-0.4*\y+5*\y);
\draw[orange] (5.8*\y+7*\y,0*\y+5*\y) -- (7*\y+7*\y,-0.4*\y+5*\y);
\draw[red] (7.3*\y+7*\y,-1.8*\y+5*\y) -- (7.3*\y+7*\y,1.5*\y+5*\y);

\filldraw[orange] (6.3*\y+7*\y,0*\y+5*\y)  node[anchor=west] {\tiny{$k$}};
\filldraw[black] (6.1*\y+7*\y,1*\y+5*\y)  node[anchor=west] {\tiny{$a$}};
\filldraw[black] (6.1*\y+7*\y,-1*\y+5*\y)  node[anchor=west] {\tiny{$b$}};
\filldraw[black] (6.7*\y+7*\y,0.7*\y+5*\y)  node[anchor=west] {\tiny{$d$}};
\filldraw[black] (6.7*\y+7*\y,-1*\y+5*\y)  node[anchor=west] {\tiny{$c$}};

\filldraw[black] (13.4*\y,0.7*\y+2*\y)  node[anchor=west] {\tiny{$u$-channel}};

\draw[] (5.5*\y+7*\y,8.5*\y) -- (6*\y+7*\y,8*\y);
\draw[] (5.5*\y+7*\y,6.5*\y) -- (6*\y+7*\y,7*\y);
\draw[orange] (6*\y+7*\y,8*\y) -- (6*\y+7*\y,7*\y);
\draw[] (6*\y+7*\y,8*\y) -- (6.5*\y+7*\y,7.7*\y);
\draw[] (6.5*\y+7*\y,7.3*\y) -- (6*\y+7*\y,7*\y);

\filldraw[black] (5.5*\y+7*\y,8.5*\y)  node[anchor=west] {\tiny{$a$}};
\filldraw[black] (5.5*\y+7*\y,6.5*\y)  node[anchor=west] {\tiny{$b$}};
\filldraw[black] (6.4*\y+7*\y,7.2*\y)  node[anchor=west] {\tiny{$c$}};
\filldraw[black] (6.4*\y+7*\y,7.8*\y)  node[anchor=west] {\tiny{$d$}};
\filldraw[orange] (5.5*\y+7*\y,7.5*\y)  node[anchor=west] {\tiny{$k$}};

\draw[<->] (9.2*\y,5*\y) -- (11.2*\y,5*\y);

\end{tikzpicture}
\end{center}
\caption{Simultaneous poles connected by a type II flip.}
\label{2_to_2_scattering_tree_level_simultaneous_poles_t_and_u_channels}
\end{figure}

Simply-laced affine Toda models comprise $r$ bosonic scalar fields $\phi_1, \dots, \phi_r$, in 1+1 dimensions, interacting through a Lagrangian
\begin{equation}
\label{Toda_theory_lagrangian_defined_in_terms_of_roots}
\mathcal{L}=\frac{1}{2} \partial_\nu \phi_a  \partial^\nu \phi_a - \frac{\mu^2}{\beta^2} \sum_{i=0}^r n_i e^{\beta \alpha^a_i  \phi_a},
\end{equation}
where $\{\alpha_i \}_{i=1}^r$ is a set of simple roots belonging to a simply laced Dynkin diagram, and $\alpha_0$ is the corresponding lowest root. Here $r$ is the rank of the associated Lie algebra and we choose to normalize all roots to have length $\sqrt{2}$. The real numbers $\mu$ and $\beta$ set the mass and the interaction scales of the model, while the integers $\{n_i\}_{i=1}^r$, called the Kac labels of the algebra, are such that with $n_0=1$ we have
$$
\sum_{i=0}^r n_i \alpha_i =0.
$$
After having  diagonalised the mass matrix coming from the second-order expansion of the potential in \eqref{Toda_theory_lagrangian_defined_in_terms_of_roots}, all non-zero $3$-point couplings respect the following area rule \cite{Braden:1989bu}
\begin{equation}
\label{Connection_among_three_point_couplings_and_areas}
C_{abc}= f_{abc} \Delta_{abc} \hspace{3mm} \text{with} \hspace{3mm} f_{abc}=\pm \frac{4\beta}{\sqrt{h}}.
\end{equation}
A universal proof of this relation was given in~\cite{Fring:1991me}, building on previous results of Freeman~\cite{Freeman:1991xw}.
In~\eqref{Connection_among_three_point_couplings_and_areas} $\Delta_{abc}$ is the area of the triangle with sides the masses of the fusing particles $a$, $b$ and $c$. The positive integer $h$ depends on the Lie algebra considered and is called the Coxeter number. The $4$- and higher-point couplings can then
be found by expanding the potential (\ref{Toda_theory_lagrangian_defined_in_terms_of_roots}) to higher orders. They satisfy various relations which allow them to be fixed in terms of 
the masses and $3$-point couplings  \cite{Fring:1992tt}, relations which turn out also to be necessary conditions for the tree-level integrability of the models
\cite{Gabai:2018tmm,Patrick_Davide_paper}. The sign, plus or minus, entering~\eqref{Connection_among_three_point_couplings_and_areas} is not the same for all the $3$-point couplings. 
The different signs, one for each non-zero $3$-point coupling, depend on the structure constants of the underlying Lie algebra and respect particular relations that prevent the presence of non-diagonal two to two processes. 
These relations emerge from the constraints of tree level integrability as we now explain, following the discussion in \cite{Patrick_Davide_paper}.
Let us consider the following process at tree level
\begin{equation}
\label{two_to_two_forbidden_process_tree_level}
a(p_1)+b(p_2) \to c(p_3)+d(p_4) ,
\end{equation}
in which we start with two initial particles of types $a$, $b$ and momenta $p_1$, $p_2$ and we finish with two outgoing particles of different types $\{c,d\} \ne \{a,b\}$.
We define the Mandelstam variables in the usual way as
\begin{equation}
\label{Definition_of_the_Mandelstam_variables_s_t_u}
s=(p_1+p_2)^2 \hspace{4mm}, \hspace{4mm} t=(p_1-p_3)^2 \hspace{4mm}, \hspace{4mm} u=(p_1-p_4)^2 .
\end{equation}
If the scattering is kinematically allowed but forbidden by integrability, all potential poles in the amplitude must cancel. In simply-laced affine Toda theories, these singularities, coming from the propagation of intermediate bound states, cancel in pairs.
This means that any time one Feynman diagram diverges due to the presence of an on-shell propagating particle, we find another diagram containing another on-shell propagator cancelling that singularity. 
The resulting connection between Feynman diagrams that are singular for the same choice of external momenta is known as the `flipping rule' \cite{Braden:1990wx}. In \cite{Patrick_Davide_paper} three different types of flip were distinguished, depending on whether the cancellation is realised between a pair of diagrams with particles propagating in the $s$- and $t$-channels, in the $s$- and $u$-channels or in the $t$- and $u$-channels. In the present paper we consider the last situation, called a type-II flip in \cite{Braden:1990wx} and \cite{Patrick_Davide_paper}.
We consider the particular situation in which the poles cancelling each other are 
due to two particles,  say $j$ and $k$, propagating in the $t$- and $u$-channels respectively\footnote{We distinguish between $t$- and $u$-channel assuming that the dual description of the former is a convex quadrilateral while the dual description of the latter is concave (see figure \ref{2_to_2_scattering_tree_level_simultaneous_poles_t_and_u_channels}).}. 
A simplification arising from working in two dimensions is that only one of the three Mandelstam variables is independent.
Therefore if in terms of the Mandelstam variable $s$ the potential
pole is at $s=s_0$, in such a position 
we have $t(s_0)=m_j^2$ and $u(s_0)=m^2_k$. 
Expanding the Mandelstam variables around the pole position $s=s_0$ we have
\begin{equation}
\begin{split}
t-m^2_j &= \frac{dt}{ds}\Bigl|_{s=s_0} (s-s_0)+\frac{1}{2}\frac{d^2t}{ds^2}\Bigl|_{s=s_0} (s-s_0)^2+\ldots \\
u-m_k^2&=\frac{du}{ds}\Bigl|_{s=s_0} (s-s_0)+\frac{1}{2}\frac{d^2u}{ds^2}\Bigl|_{s=s_0} (s-s_0)^2+\ldots ,
\end{split}
\end{equation}
so that the sum of the corresponding divergent tree-level Feynman diagrams is
\begin{multline}
\label{tree_level_sum_of_two_channels_singular_part_plus_finite_contribution}
\frac{C_{a\bar{c}\bar{j}} C_{j c \bar{d}}}{t-m^2_j}+ \frac{C_{a \bar{d} \bar{k}}C_{k b \bar{c}}}{u-m^2_k} =\\
\frac{1}{s-s_0} \biggl( \frac{C_{a\bar{c}\bar{j}} C_{j c \bar{d}}}{\frac{dt}{ds}\bigl|_{s_0} }+ \frac{C_{a \bar{d} \bar{k}}C_{k b \bar{c}}}{\frac{du}{ds}\bigl|_{s_0}}\biggr)-\frac{1}{2} \Bigl( C_{a\bar{c}\bar{j}} C_{j c \bar{d}} \frac{\frac{d^2t}{ds^2}\bigl|_{s_0}}{\bigl(\frac{dt}{ds}\bigl|_{s_0}\bigr)^2} + C_{a \bar{d} \bar{k}}C_{k b \bar{c}} \frac{\frac{d^2u}{ds^2}\bigl|_{s_0}}{\bigl(\frac{du}{ds}\bigl|_{s_0}\bigr)^2} \Bigr).
\end{multline}
Terms having positive powers of $s-s_0$ have been omitted in the expression above since they 
vanish at $s=s_0$.

Poles corresponding to stable bound states lie on the physical strip,
and are purely imaginary. After a possible overall Lorentz boost, this means that the rapidities $\theta$ of the interacting particles can also be taken to be purely imaginary, 
with $\theta=i U$ and $U$ real. 
The on-shell particle momenta $p=m (\cosh \theta, \sinh \theta)=m (\cos U, i \sin U)$ can be represented as complex numbers $m(\cos U+i\sin U)$ whose absolute values the masses of the particles and whose phases are given by the real numbers $U$, with
the plane of real energies and purely imaginary spatial momenta in which they lie inheriting a Euclidean metric. The
duals of Feynman diagrams for two to two processes can also be drawn in this plane, as tiled quadrilaterals whose external sides have lengths equal to the masses of the interacting particles. 
For the particular process we are studying, such a description at the value $s=s_0$ corresponding to the potential pole is shown
in figure \ref{2_to_2_scattering_tree_level_simultaneous_poles_t_and_u_channels}, where 
the lengths of the red, blue and orange diagonals correspond to $\sqrt{s_0}$, $m_j$ and $m_k$ respectively. This illustrates the fact that in disallowed $4$-point processes such as this one, the flipping rule translates into the statement that any time one diagonal is on-shell (i.e.\ its length is equal to the mass of the associated propagating particle), corresponding to a singular propagator in one of the $s$-, $t$- or $u$-channels, then for the same values of the external kinematics exactly one other on-shell diagonal is on-shell, 
so that the sum of the two diagrams can be finite. This re-tiling property
can be proved in a universal fashion by considering alternative projections of tetrahedra in root space \cite{Dorey:1990xa}. 

The flipping rule means that it is possible for 
the coefficient of $(s-s_0)^{-1}$ in \eqref{tree_level_sum_of_two_channels_singular_part_plus_finite_contribution} to be zero; that it does actually vanish can be
proven using the following geometrical identities
\begin{equation}
\label{derivatives_of_s_respect_to_t_and_u}
\frac{dt}{ds}\Bigl|_{s_0}=- \frac{\Delta_{acj} \Delta_{bdj}}{\Delta_{abi} \Delta_{cdi}} \hspace{6mm}, \hspace{6mm} \frac{du}{ds}\Bigl|_{s_0}= \frac{\Delta_{adk} \Delta_{bck}}{\Delta_{abi} \Delta_{cdi}}
\end{equation}
connecting the derivatives of the squared diagonals (performed keeping the lengths of the external sides $\{a, b, c, d\}$ fixed at the values $\{m_a, m_b, m_c, m_d\}$) to the areas of the fusing mass triangles  \cite{Patrick_Davide_paper}. 
Substituting this expression into \eqref{tree_level_sum_of_two_channels_singular_part_plus_finite_contribution} and using the area rule \eqref{Connection_among_three_point_couplings_and_areas} we find that the singular part in~\eqref{tree_level_sum_of_two_channels_singular_part_plus_finite_contribution} is proportional to
$$
-f_{a\bar{c}\bar{j}} f_{j c \bar{d}}+f_{a\bar{d}\bar{k}} f_{k b \bar{c}}.
$$
This expression is indeed zero for all the pairs of Feynman diagrams connected by a type II flip. This follows from  properties of the structure constants of the Lie algebra associated with the model; a general proof is given in \cite{Patrick_Davide_paper}. This is a fundamental requirement for the cancellation of the singularity in the amplitude.
We can therefore say that type II flips preserve the sign of the product of the $f$-functions entering in $3$-point couplings.

On the other hand, the finite term in \eqref{tree_level_sum_of_two_channels_singular_part_plus_finite_contribution} is not zero but is instead equal to minus the sum of all the remaining tree-level Feynman diagrams evaluated at $s=s_0$, in such a way that after having summed over all the Feynman diagrams the result is zero. This is an interesting fact that must be true given that the total amplitude at the tree level is zero, independently of the choice of the external kinematics \cite{Patrick_Davide_paper}: the geometry of the singular graphs depicted in figure~\ref{2_to_2_scattering_tree_level_simultaneous_poles_t_and_u_channels} also
encodes  the value of the sum of all the remaining Feynman diagrams. This fact will be of great importance at the moment we will break loop integrals into products of tree-level diagrams to compute the residues at the poles, as we will discuss in the next section.

\section{Singularities from Feynman diagram cuts}
\label{Singularities_from_Feynman_diagrams_main_section}
Higher-order poles in two dimensional S-matrices have an interpretation in terms of Landau (or anomalous threshold) singularities according to the Coleman-Thun mechanism \cite{Coleman:1978kk}. For particular values of the external momenta, it is possible that multiple internal propagators, at a particular value of the loop integration variable $l$, go  on-shell simultaneously, generating divergences. In \cite{Braden:1990wx} a technique to obtain the residues of such singularities was explained, through which second- and third-order Landau singularities were computed for the ADE series of models. In this section we show how to calculate the singular part of one-loop Feynman diagrams in an alternative way.
The purpose is to show that on the pole position loop diagrams are nothing but products of one or more different tree-level graphs in which on-shell bound states propagate. How the loop can be cut into such tree-level graphs is determined by its on-shell geometry.

\subsection{Computing the box integral}
\label{subsection_explanation_of_the_method}
We explain the adopted method with a well-known example, the box integral. 
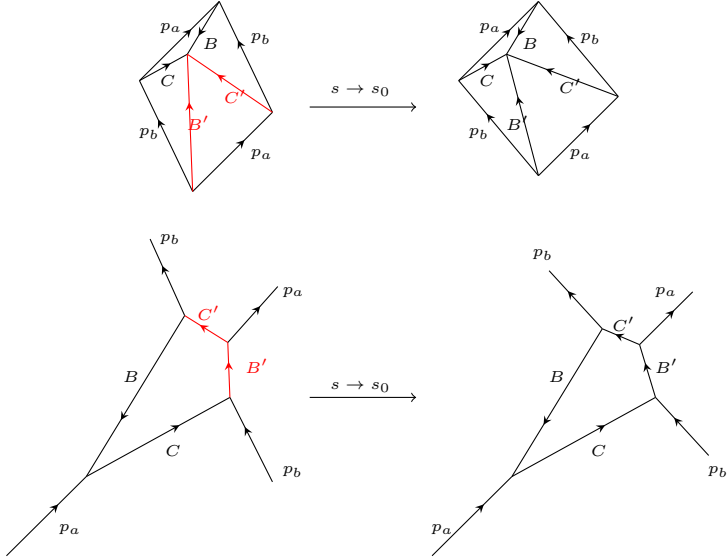
\begin{figure}
\medskip
\begin{center}
\begin{tikzpicture}
\tikzmath{\y = 0.7;}

%Deformed On-shell box diagram      
\draw[directed] (5.8*\y,1*\y) -- (7.3*\y,2.5*\y);
\draw[directed] (5.8*\y,1*\y) -- (6.7*\y,1.5*\y);
\draw[directed] (7.3*\y,2.5*\y) -- (6.7*\y,1.5*\y);
\draw[directed] (6.8*\y,-1.1*\y) -- (5.8*\y,1*\y);
\draw[directed] (6.8*\y,-1.1*\y) -- (8.3*\y,0.4*\y);
\draw[directed] (8.3*\y,0.4*\y) -- (7.3*\y,2.5*\y);
\draw[directed][red] (6.8*\y,-1.1*\y) -- (6.7*\y,1.5*\y);
\draw[directed][red] (8.3*\y,0.4*\y) -- (6.7*\y,1.5*\y);

\filldraw[black] (6*\y,2*\y)  node[anchor=west] {\tiny{$p_a$}};
\filldraw[black] (5.6*\y,0*\y)  node[anchor=west] {\tiny{$p_b$}};
\filldraw[black] (7.7*\y,-0.4*\y)  node[anchor=west] {\tiny{$p_a$}};
\filldraw[black] (7.7*\y,1.8*\y)  node[anchor=west] {\tiny{$p_b$}};
\filldraw[black] (6*\y,1*\y)  node[anchor=west] {\tiny{$C$}};
\filldraw[black] (6.8*\y,1.7*\y)  node[anchor=west] {\tiny{$B$}};
\filldraw[red] (6.5*\y,0.2*\y)  node[anchor=west] {\tiny{$B'$}};
\filldraw[red] (7.2*\y,0.7*\y)  node[anchor=west] {\tiny{$C'$}};

\draw [->] (9*\y,0.5*\y) -- (11*\y,0.5*\y);
\filldraw[black] (9.2*\y,0.8*\y)  node[anchor=west] {\tiny{$s \to s_0$}};

%On-shell box diagram on the pole    
\draw[directed] (5.8*\y+6*\y,1*\y) -- (7.3*\y+6*\y,2.5*\y);
\draw[directed] (5.8*\y+6*\y,1*\y) -- (6.7*\y+6*\y,1.5*\y);
\draw[directed] (7.3*\y+6*\y,2.5*\y) -- (6.7*\y+6*\y,1.5*\y);
\draw[directed] (7.3*\y+6*\y,-0.8*\y) -- (5.8*\y+6*\y,1*\y);
\draw[directed] (7.3*\y+6*\y,-0.8*\y) -- (8.8*\y+6*\y,0.7*\y);
\draw[directed] (8.8*\y+6*\y,0.7*\y) -- (7.3*\y+6*\y,2.5*\y);
\draw[directed][] (8.8*\y+6*\y, 0.7*\y) -- (6.7*\y+6*\y,1.5*\y);
\draw[directed][] (7.3*\y+6*\y,-0.8*\y) -- (6.7*\y+6*\y,1.5*\y);

\filldraw[black] (6*\y+6*\y,2*\y)  node[anchor=west] {\tiny{$p_a$}};
\filldraw[black] (5.8*\y+6*\y,0*\y)  node[anchor=west] {\tiny{$p_b$}};
\filldraw[black] (7.7*\y+6*\y,-0.4*\y)  node[anchor=west] {\tiny{$p_a$}};
\filldraw[black] (7.9*\y+6*\y,1.8*\y)  node[anchor=west] {\tiny{$p_b$}};
\filldraw[black] (6*\y+6*\y,1*\y)  node[anchor=west] {\tiny{$C$}};
\filldraw[black] (6.8*\y+6*\y,1.7*\y)  node[anchor=west] {\tiny{$B$}};
\filldraw[] (6.5*\y+6*\y,0.2*\y)  node[anchor=west] {\tiny{$B'$}};
\filldraw[] (7.5*\y+6*\y,0.9*\y)  node[anchor=west] {\tiny{$C'$}};

%Deformed On-shell Feynman diagram       
\draw[directed] (4.8*\y-1.5*\y,-6*\y-2*\y) -- (6.3*\y-1.5*\y,-4.5*\y-2*\y);
\draw[directed] (6.3*\y-1.5*\y,-4.5*\y-2*\y) -- (6.3*\y+2.7*\y-1.5*\y,-4.5*\y+1.5*\y-2*\y);
\draw[directed] (8.15*\y-1.5*\y,-1.45*\y-2*\y) -- (6.3*\y-1.5*\y,-4.5*\y-2*\y);
\draw[directed] (8.3*\y,-6.6*\y) -- (7.5*\y,-5*\y);
\draw[directed][red] (9*\y-1.5*\y,-3*\y-2*\y) -- (9*\y-0.04*\y-1.5*\y,-3*\y+1.04*\y-2*\y);
\draw[directed][red] (9*\y-0.04*\y-1.5*\y,-2*\y+0.04*\y-2*\y) -- (8.15*\y-1.5*\y,-1.45*\y-2*\y);
\draw[directed] (9*\y-0.04*\y-1.5*\y,-2*\y+0.04*\y-2*\y) -- (8.4*\y,-2.9*\y);
\draw[directed] (8.15*\y-1.5*\y,-1.45*\y-2*\y) -- (6*\y,-2*\y);

\filldraw[black] (6*\y,-2*\y)  node[anchor=west] {\tiny{$p_b$}};
\filldraw[black] (8.3*\y,-3*\y)  node[anchor=west] {\tiny{$p_a$}};
\filldraw[black] (8.3*\y,-6.4*\y)  node[anchor=west] {\tiny{$p_b$}};
\filldraw[black] (4.1*\y,-7.5*\y)  node[anchor=west] {\tiny{$p_a$}};
\filldraw[black] (6.1*\y,-6*\y)  node[anchor=west] {\tiny{$C$}};
\filldraw[red] (7.6*\y,-4.4*\y)  node[anchor=west] {\tiny{$B'$}};
\filldraw[black] (5.3*\y,-4.6*\y)  node[anchor=west] {\tiny{$B$}};
\filldraw[red] (6.7*\y,-3.4*\y)  node[anchor=west] {\tiny{$C'$}};

\draw [->] (9*\y,-5*\y) -- (11*\y,-5*\y);
\filldraw[black] (9.2*\y,-4.8*\y)  node[anchor=west] {\tiny{$s \to s_0$}};

%On-shell Feynman diagram      
\draw[directed] (4.8*\y+6.5*\y,-6*\y-2*\y) -- (6.3*\y+6.5*\y,-4.5*\y-2*\y);
\draw[directed] (6.3*\y+6.5*\y,-4.5*\y-2*\y) -- (6.3*\y+2.7*\y+6.5*\y,-4.5*\y+1.5*\y-2*\y);
\draw[directed] (14.5*\y,-3.7*\y) -- (6.3*\y+6.5*\y,-4.5*\y-2*\y);
\draw[directed] (16.5*\y,-6.1*\y) -- (15.5*\y,-5*\y);
\draw[directed][] (15.5*\y,-5*\y) -- (15.2*\y,-4*\y);
\draw[directed][] (15.2*\y,-4*\y) -- (14.5*\y,-3.7*\y);
\draw[directed] (14.5*\y,-3.7*\y) -- (13.5*\y,-2.6*\y);
\draw[directed] (15.2*\y,-4*\y) -- (16.2*\y,-3*\y);

\filldraw[black] (6*\y+7*\y,-2.3*\y)  node[anchor=west] {\tiny{$p_b$}};
\filldraw[black] (8.3*\y+7*\y,-3*\y)  node[anchor=west] {\tiny{$p_a$}};
\filldraw[black] (9.3*\y+7*\y,-6.4*\y)  node[anchor=west] {\tiny{$p_b$}};
\filldraw[black] (4.1*\y+7*\y,-7.5*\y)  node[anchor=west] {\tiny{$p_a$}};
\filldraw[black] (7.1*\y+7*\y,-6*\y)  node[anchor=west] {\tiny{$C$}};
\filldraw[] (9.3*\y+6*\y,-4.4*\y)  node[anchor=west] {\tiny{$B'$}};
\filldraw[black] (7.3*\y+6*\y,-4.6*\y)  node[anchor=west] {\tiny{$B$}};
\filldraw[] (8.5*\y+6*\y,-3.6*\y)  node[anchor=west] {\tiny{$C'$}};

\end{tikzpicture}
\end{center}

\caption{Box diagram contributing to the pole (bottom row) and its dual description (top row). The loop integration variable $l$ is chosen so that  $B^2=m^2_B$ and $C^2=m^2_C$ when $l=(0,0)$. For general external kinematics
(the left hand column) the momenta $B'$ and $C'$ are off-shell at this point; they become on-shell when $s=s_0$, the pole position (the right hand column). On-shell momenta are coloured black, off-shell red.}
\label{Example_of_on_shell_one_loop_box_diagram_how_to_perform_the_limit}
\end{figure}
Referring to figure \ref{Example_of_on_shell_one_loop_box_diagram_how_to_perform_the_limit} we consider the scattering of two bosonic  particles $a$ and $b$ near a particular value $s_0$ of the Mandelstam variable $s=(p_a+p_b)^2$ at which there exists a point in the loop integration region where all the internal propagators $B$, $C$, $B'$ and $C'$ are on-shell at the same time. With a small abuse of notation we label in capital letters both the types of particles propagating inside the box and their momenta.
Since the loop carries two degrees of freedom it is possible to arrange for two internal propagators to be on-shell when the loop integration variable $l=(0,0)$. 
In our case, we fix
\begin{equation}
\label{onshell_condition_for_B_and_C_in_the_integral}
B^2-m^2_B=0 \ \ \text{and} \ \ C^2-m^2_C=0.
\end{equation}
We will comment on this in  more detail in one moment.
For general external kinematics the remaining two propagators, corresponding to the particles $B'$ and $C'$, are off-shell at this point and only become on-shell in the limit $s\to s_0$, as shown on the RHS of figure \ref{Example_of_on_shell_one_loop_box_diagram_how_to_perform_the_limit}. In the top row of figure \ref{Example_of_on_shell_one_loop_box_diagram_how_to_perform_the_limit} we see that the lengths of the two black diagonals ($B$ and $C$) are fixed as $s$ varies, maintaining their on-shell values. On the other hand, away from $s=s_0$, $B'$ and $C'$, coloured red, have lengths different from their masses and assume these values only when $s=s_0$, i.e.\ at the singularity. 
In the neighbourhood of $s=s_0$, we can expand $B'^2$ and $C'^2$ in terms of $s-s_0$ as
\begin{equation}
\label{expansion_up_to_ssquare_of_Bprime_and_Cprime}
\begin{split}
B'^2-m^2_{B'}&= \frac{d B'^2}{ds}\Bigr|_{B,C} (s-s_0)+\frac{1}{2} \frac{d^2 B'^2}{ds^2}\Bigr|_{B,C} (s-s_0)^2 + \ldots\\
C'^2-m^2_{B'}&=\frac{d C'^2}{ds}\Bigr|_{B,C} (s-s_0)+\frac{1}{2} \frac{d^2 C'^2}{ds^2}\Bigr|_{B,C} (s-s_0)^2 + \ldots .
\end{split}
\end{equation}
The subscripts $B$ and $C$ in \eqref{expansion_up_to_ssquare_of_Bprime_and_Cprime} mean that we are taking the derivatives with the squared momenta $B^2=m^2_B$ and $C^2=m^2_C$ held fixed.
Using the relations in  \eqref{onshell_condition_for_B_and_C_in_the_integral}, \eqref{expansion_up_to_ssquare_of_Bprime_and_Cprime} and scaling $l=(s-s_0) \tilde{l}$ we see that the  integral near the pole can be written as
\begin{equation}
\label{Expression_for_I1_intitial_step_in_computing_the_box}
\begin{split}
I&=\int \frac{d^2 l}{(2 \pi)^2} \frac{1}{(B + l)^2 -m^2_B+i \epsilon} \ \frac{1}{(C + l)^2 -m^2_C+i \epsilon} \\
&\times \frac{1}{(B' + l)^2 -m^2_{B'}+i \epsilon} \ \frac{1}{(C' + l)^2 -m^2_{C'}+i \epsilon}\\
&\sim\frac{1}{(s-s_0)^2}\int \frac{d^2 \tilde{l}}{(2 \pi)^2} \frac{1}{2 B \cdot \tilde{l} + (s-s_0)\tilde{l}^2+i \epsilon} \ \frac{1}{2 C \cdot \tilde{l} + (s-s_0)\tilde{l}^2+i \epsilon} \\
&\times    \frac{1}{\frac{d B'^2}{ds} + 2 B' \cdot \tilde{l}+\frac{1}{2} \frac{d^2 B'^2}{ds^2} (s-s_0) + (s-s_0) \tilde{l}^2+i \epsilon}\\
&\times \frac{1}{\frac{d C'^2}{ds} + 2 C' \cdot \tilde{l}+\frac{1}{2} \frac{d^2 C'^2}{ds^2} (s-s_0) + (s-s_0) \tilde{l}^2+i \epsilon}
\end{split}
\end{equation}
where we have omitted the factors coming from the $3$-point couplings, and dropped second and higher powers of $(s-s_0)$ from the last two denominators in the second expression.

\begin{figure}
\medskip
\begin{center}
\begin{tikzpicture}
\tikzmath{\y = 0.7;}

\draw[->][orange] (11.8*\y,1*\y) -- (11.8*\y+4.5*\y,1*\y+2.5*\y);
\draw[dashed][orange] (11.8*\y,1*\y) -- (11.8*\y-1.8*\y,1*\y-1*\y);
\draw[->][orange] (11.8*\y,1*\y) -- (11.8*\y-2.5*\y,1*\y+4.5*\y);
\draw[dashed][orange] (11.8*\y,1*\y) -- (11.8*\y+1*\y,1*\y-1.8*\y);
\draw[->][] (11.8*\y,1*\y) -- (11.8*\y-2.5*\y,1*\y-4.5*\y);
\draw[directed][orange] (8.8*\y+6*\y, 0.7*\y) -- (6.7*\y+6*\y,1.5*\y);
\draw[directed][orange] (7.3*\y+6*\y,-0.8*\y) -- (6.7*\y+6*\y,1.5*\y);
\filldraw[orange] (9.9*\y+6*\y,2.9*\y)  node[anchor=west] {\small{$\Re(E)$}};
\filldraw[orange] (3.5*\y+6*\y,5.7*\y)  node[anchor=west] {\small{$\Im(P)$}};
\filldraw[] (3.8*\y+6*\y,-3*\y)  node[anchor=west] {\small{$\Re(P)$}};

\draw[directed][orange] (5.8*\y+6*\y,1*\y) -- (7.3*\y+6*\y,2.5*\y);
\draw[directed][orange] (11.8*\y,1*\y) -- (11.8*\y+0.9*\y,1*\y+0.5*\y);
\draw[directed][orange] (7.3*\y+6*\y,2.5*\y) -- (6.7*\y+6*\y,1.5*\y);
\draw[directed][orange] (7.3*\y+6*\y,-0.8*\y) -- (5.8*\y+6*\y,1*\y);
\draw[directed][orange] (7.3*\y+6*\y,-0.8*\y) -- (8.8*\y+6*\y,0.7*\y);
\draw[directed][orange] (8.8*\y+6*\y,0.7*\y) -- (7.3*\y+6*\y,2.5*\y);

\filldraw[orange] (6*\y+6*\y,2*\y)  node[anchor=west] {\tiny{$p_a$}};
\filldraw[orange] (5.8*\y+6*\y,0*\y)  node[anchor=west] {\tiny{$p_b$}};
\filldraw[orange] (7.7*\y+6*\y,-0.4*\y)  node[anchor=west] {\tiny{$p_a$}};
\filldraw[orange] (7.9*\y+6*\y,1.8*\y)  node[anchor=west] {\tiny{$p_b$}};
\filldraw[orange] (6*\y+6*\y,1*\y)  node[anchor=west] {\tiny{$C$}};
\filldraw[orange] (6.8*\y+6*\y,1.7*\y)  node[anchor=west] {\tiny{$B$}};
\filldraw[orange] (6.5*\y+6*\y,0.2*\y)  node[anchor=west] {\tiny{$B'$}};
\filldraw[orange] (7.5*\y+6*\y,0.9*\y)  node[anchor=west] {\tiny{$C'$}};

\end{tikzpicture}
\end{center}

\caption{The box diagram lying in the Euclidean slice of the space of complex energy and momentum.}
\label{Figure_showing_the_integration_region}
\end{figure}
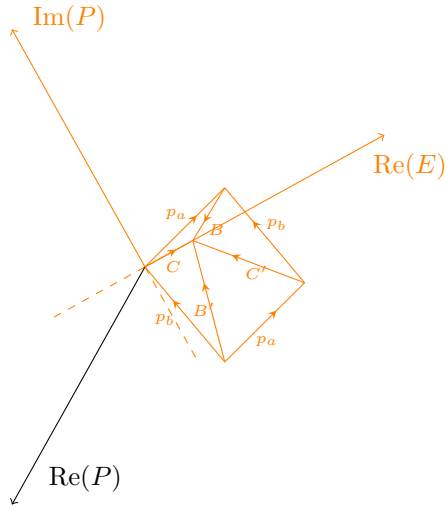

To have a better idea of the surface we are integrating over we should highlight that, just as discussed following equation \eqref{tree_level_sum_of_two_channels_singular_part_plus_finite_contribution} in the context of tree-level diagrams, the external momenta at which the box diagram yields a pole can be taken to have purely imaginary rapidities, and the same is true of all internal momenta which go on-shell at the singularity. This implies that they lie in the Euclidean plane of real energies and imaginary momenta, on which Lorentz boosts act as simple rotations.
It is therefore not restrictive to depict the dual configuration as in figure~\ref{Figure_showing_the_integration_region}, where we choose the axis of real energy to be aligned with $C$. 
The on-shell box is then contained in the orange plane of figure~\ref{Figure_showing_the_integration_region}. The integration variable $l_0$ takes values on the axis $\Re(E)$ and has to be performed between $-\infty$ and $+\infty$. 
As discussed in appendix~\ref{Appendix_contour_integrals}, for these values of the external momenta the variable $l_1$ should instead to take values on the imaginary momentum axis $\Im(P)$ and be integrated between $+i\infty$ and $-i \infty$. This means that the integration over $l$ has to be performed in the same plane where the box lives and we can always translate $l$ in this plane so that, at the point $l=(0,0)$, $B$ and $C$ are on-shell.
If we choose the centre of the space as in figure~\ref{Figure_showing_the_integration_region}, this configuration is reached at the value $l_0=m_C$ and $l_1=0$.
We can therefore translate $l_0$ along the energy axis so to have the pole at the position $l_0=l_1=0$, as assumed in~\eqref{onshell_condition_for_B_and_C_in_the_integral}. 

Continuing our computation, in~\eqref{Expression_for_I1_intitial_step_in_computing_the_box} the residue is isolated and we can simply compute the leading contribution by setting $s=s_0$ in all the terms apart from the overall factor $(s-s_0)^{-2}$. 
If we expand the integral in powers of $(s-s_0)$ the singular part of~\eqref{Expression_for_I1_intitial_step_in_computing_the_box} can be split into two terms, one ($I_2$) presenting a second order singularity of the form $(s-s_0)^{-2}$, the other ($I_1$) presenting a simple pole in $(s-s_0)$
\begin{equation}
\label{singular_part_of_I_divided_in_leading_and_subleading}
    I^{\text{(sing)}}=I_2+I_1.
\end{equation}
Here and elsewhere, for a  function $f$ with Laurent series
$$
f(s)=\sum_{p=-\infty}^{P} \frac{f_p}{(s-s_0)^p} 
$$
around a pole $s_0$ of order $P$, we define its singular part as
\begin{equation}
\label{definition_of_singular_part}
    f(s)^{(\text{sing})}=\sum_{p=1}^{P} \frac{f_p}{(s-s_0)^p}.
\end{equation}
The same definition was used in~\eqref{Expansion_of_S_around_an_arbitrary_2N_order_pole} to give the S-matrix expansion around higher-order poles.

The leading order term in~\eqref{singular_part_of_I_divided_in_leading_and_subleading} is given by
\begin{equation}
\label{integral_expression_for_I2_first_integral}
I_2=\frac{1}{(s-s_0)^2}\int \frac{d^2 \tilde{l}}{(2 \pi)^2} \frac{1}{2 B \cdot \tilde{l} +i \epsilon} \frac{1}{2 C \cdot \tilde{l} +i \epsilon} \ \frac{1}{\frac{d B'^2}{ds}\bigr|_{B,C} +2 B' \cdot \tilde{l} +i \epsilon} \frac{1}{\frac{d C'^2}{ds}\bigr|_{B,C}  +2 C' \cdot \tilde{l} +i \epsilon}.
\end{equation}
A feature of our choice of on-shell momenta ($B$ and $C$), around which we are expanding the integral, is that each other momentum can be expressed as a negative linear combination of them
\begin{equation}
\label{Bprime_and_Cprime_as_linear_combinations_of_B_and_C}
B'=-\frac{\Delta_{B'C}}{\Delta_{BC}} B  -\frac{\Delta_{B'B}}{\Delta_{BC}}C \hspace{4mm} \text{and} \hspace{4mm} C'=-\frac{\Delta_{C'C}}{\Delta_{BC}} B  -\frac{\Delta_{C'B}}{\Delta_{BC}}C.
\end{equation}
In the expressions above $\Delta_{XY}$ indicates the area of a triangle having for sides the two vectors $X$ and $Y$. (Since two vectors are sufficient to identify a triangle uniquely we can
omit the third side of the triangle; for example, referring to figure \ref{Example_of_on_shell_one_loop_box_diagram_how_to_perform_the_limit}, we just write $\Delta_{B'C}$ instead of $\Delta_{B'Cb}$.)
The relations in \eqref{Bprime_and_Cprime_as_linear_combinations_of_B_and_C} are not a surprise if we look at the on-shell Feynman diagram in figure \ref{Example_of_on_shell_one_loop_box_diagram_how_to_perform_the_limit}. We note indeed that both the vector $B'$ and $C'$ (the red arrows in the figure) belong to the region of the plane spanned by $-B$ and $-C$. The fact that all the coefficients in \eqref{Bprime_and_Cprime_as_linear_combinations_of_B_and_C} are negative makes the integration particularly simple, as we will see in one moment.

Changing integration variables
\begin{equation}
\label{u_and_v_as_a_function_of_tilde_l}
2 B \cdot \tilde{l} = u \hspace{4mm} \text{and} \hspace{4mm} 2 C \cdot \tilde{l} = v
\end{equation}
and taking into account the Jacobian of the transformation
\begin{equation}
\label{jacobian_of_the_change_of_variables_box_integral_example}
d^2 \tilde{l}=\frac{du \ dv}{8 i \Delta_{BC}}
\end{equation}
the integral can be written as
\begin{equation}
\begin{split}
I_2&=\frac{1}{(s-s_0)^2} \int \frac{du\ dv}{(2 \pi)^2 8 i \Delta_{BC}} \frac{1}{u+i \epsilon} \frac{1}{v+i \epsilon}\\
&\times \frac{1}{\frac{d B'^2}{ds}\bigr|_{B,C} -\frac{\Delta_{B'C}}{\Delta_{BC}} u  -\frac{\Delta_{B'B}}{\Delta_{BC}} v +i \epsilon} \frac{1}{\frac{d C'^2}{ds}\bigr|_{B,C} -\frac{\Delta_{C'C}}{\Delta_{BC}} u  -\frac{\Delta_{C'B}}{\Delta_{BC}}v  +i \epsilon}.
\end{split}
\end{equation}
Since $\tilde{l}_0$ was integrated between $-\infty$ and $+\infty$, and $\tilde{l}_1$ was integrated between $+i\infty$ and $-i\infty$, the variables $u$ and $v$ are both integrated on the real axis between $-\infty$ and $+\infty$.
The result can be obtained by closing both the $u$ and $v$ contours in the lower half complex plane. Since all the coefficients in \eqref{Bprime_and_Cprime_as_linear_combinations_of_B_and_C} are negative, the $B'$- and $C'$-propagators have poles on the opposite side of the real axis compared to $B$ and $C$ and using Cauchy's theorem we have 
\begin{equation}
\label{leading_term_to_the_box_integral_B_C_cut}
I_2=\frac{1}{(s-s_0)^2} \frac{i}{8 \Delta_{BC}} \frac{1}{\frac{d B'^2}{ds}} \frac{1}{\frac{d C'^2}{ds}}.
\end{equation}
We note that the area $\Delta_{BC}$ (or if we prefer $\Delta_{aBC}$) does not depend on $s$ and is fixed along the limit $s\to s_0$. This is because all three sides $p_a$, $B$ and $C$ are on-shell --  the small triangle $\Delta_{aBC}$ is the same on the LHS and  the RHS of the
top row of figure \ref{Example_of_on_shell_one_loop_box_diagram_how_to_perform_the_limit}. Similarly, at the bottom of the figure, we see that the fusing angles at the vertex $C_{aB \bar{C}}$ stay fixed along the limit. In this case, the Jacobian factor does not present any further $s$-expansion, since it is constant in $s$, and the $I_2$ term does not contain any subleading power of $(s-s_0)$. 

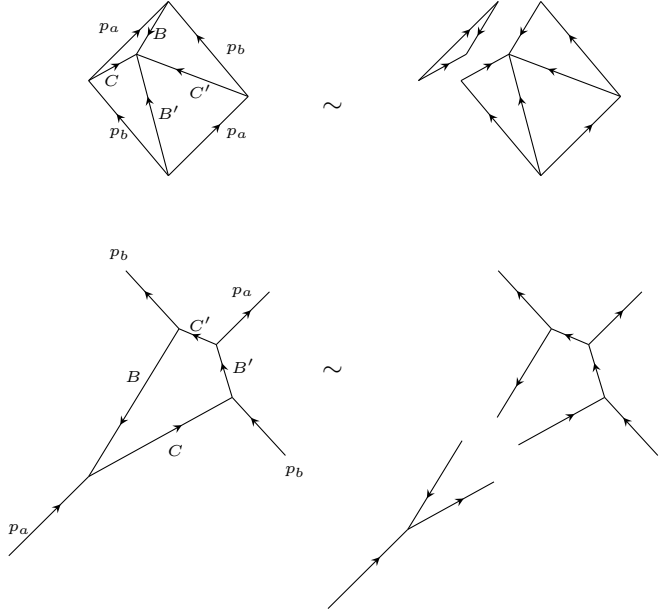
\begin{figure}
\medskip
\begin{center}
\begin{tikzpicture}
\tikzmath{\y = 0.7;}

%On-shell box diagram               
\draw[directed] (0.8*\y,1*\y-5*\y) -- (2.3*\y,2.5*\y-5*\y);
\draw[directed] (0.8*\y,1*\y-5*\y) -- (1.7*\y,1.5*\y-5*\y);
\draw[directed] (2.3*\y,2.5*\y-5*\y) -- (1.7*\y,1.5*\y-5*\y);
\draw[directed] (2.3*\y,-0.8*\y-5*\y) -- (0.8*\y,1*\y-5*\y);
\draw[directed] (2.3*\y,-0.8*\y-5*\y) -- (1.7*\y,1.5*\y-5*\y);
\draw[directed] (2.3*\y,-0.8*\y-5*\y) -- (3.8*\y,0.7*\y-5*\y);
\draw[directed] (3.8*\y,0.7*\y-5*\y) -- (2.3*\y,2.5*\y-5*\y);
\draw[directed] (3.8*\y, 0.7*\y-5*\y) -- (1.7*\y,1.5*\y-5*\y);

\filldraw[black] (0.8*\y,2*\y-5*\y)  node[anchor=west] {\tiny{$p_a$}};
\filldraw[black] (1*\y,0*\y-5*\y)  node[anchor=west] {\tiny{$p_b$}};
\filldraw[black] (3.2*\y,0*\y-5*\y)  node[anchor=west] {\tiny{$p_a$}};
\filldraw[black] (3.2*\y,1.6*\y-5*\y)  node[anchor=west] {\tiny{$p_b$}};
\filldraw[black] (0.9*\y,1*\y-5*\y)  node[anchor=west] {\tiny{$C$}};
\filldraw[black] (1.9*\y,0.4*\y-5*\y)  node[anchor=west] {\tiny{$B'$}};
\filldraw[black] (2.5*\y,0.8*\y-5*\y)  node[anchor=west] {\tiny{$C'$}};
\filldraw[black] (1.8*\y,1.9*\y-5*\y)  node[anchor=west] {\tiny{$B$}};

\filldraw[black] (5*\y,-4.5*\y)  node[anchor=west] {\small{$\sim$}};

%On-shell box diagram cut  
\draw[directed] (0.8*\y+6.2*\y,1*\y-5*\y) -- (2.3*\y+6.2*\y,2.5*\y-5*\y);
\draw[directed] (0.8*\y+6.2*\y,1*\y-5*\y) -- (1.7*\y+6.2*\y,1.5*\y-5*\y);
\draw[directed] (2.3*\y+6.2*\y,2.5*\y-5*\y) -- (1.7*\y+6.2*\y,1.5*\y-5*\y);

\draw[directed] (0.8*\y+7*\y,1*\y-5*\y) -- (1.7*\y+7*\y,1.5*\y-5*\y);
\draw[directed] (2.3*\y+7*\y,2.5*\y-5*\y) -- (1.7*\y+7*\y,1.5*\y-5*\y);
\draw[directed] (2.3*\y+7*\y,-0.8*\y-5*\y) -- (0.8*\y+7*\y,1*\y-5*\y);
\draw[directed] (2.3*\y+7*\y,-0.8*\y-5*\y) -- (1.7*\y+7*\y,1.5*\y-5*\y);
\draw[directed] (2.3*\y+7*\y,-0.8*\y-5*\y) -- (3.8*\y+7*\y,0.7*\y-5*\y);
\draw[directed] (3.8*\y+7*\y,0.7*\y-5*\y) -- (2.3*\y+7*\y,2.5*\y-5*\y);
\draw[directed] (3.8*\y+7*\y, 0.7*\y-5*\y) -- (1.7*\y+7*\y,1.5*\y-5*\y);

\filldraw[black] (5*\y,-9.5*\y)  node[anchor=west] {\small{$\sim$}};

%On-shell Feynman diagram     
\draw[directed] (-0.7*\y,-13*\y) -- (0.8*\y,-11.5*\y);
\draw[directed] (0.8*\y,-11.5*\y) -- (0.8*\y+2.7*\y,-10*\y);
\draw[directed] (2.5*\y,-8.7*\y) -- (0.8*\y,-11.5*\y);
\draw[directed] (4.5*\y,-11.1*\y) -- (3.5*\y,-10*\y);
\draw[directed][] (3.5*\y,-10*\y) -- (3.2*\y,-9*\y);
\draw[directed][] (3.2*\y,-9*\y) -- (2.5*\y,-8.7*\y);
\draw[directed] (2.5*\y,-8.7*\y) -- (1.5*\y,-7.6*\y);
\draw[directed] (3.2*\y,-9*\y) -- (4.2*\y,-8*\y);

\filldraw[black] (1*\y,-7.3*\y)  node[anchor=west] {\tiny{$p_b$}};
\filldraw[black] (3.3*\y,-8*\y)  node[anchor=west] {\tiny{$p_a$}};
\filldraw[black] (4.3*\y,-11.4*\y)  node[anchor=west] {\tiny{$p_b$}};
\filldraw[black] (-0.9*\y,-12.5*\y)  node[anchor=west] {\tiny{$p_a$}};
\filldraw[black] (2.1*\y,-11*\y)  node[anchor=west] {\tiny{$C$}};
\filldraw[] (3.3*\y,-9.4*\y)  node[anchor=west] {\tiny{$B'$}};
\filldraw[black] (1.3*\y,-9.6*\y)  node[anchor=west] {\tiny{$B$}};
\filldraw[] (2.5*\y,-8.6*\y)  node[anchor=west] {\tiny{$C'$}};

%On-shell Feynman diagram cut  

\draw[directed] (5.3*\y,-14*\y) -- (6.8*\y,-12.5*\y);
\draw[directed] (6.8*\y,-12.5*\y) -- (6.8*\y+0.6*2.7*\y,-12.5*\y+0.6*1.5*\y);
\draw[directed] (6.8*\y+0.6*1.7*\y,-12.5*\y+0.6*2.8*\y) -- (6.8*\y,-12.5*\y);
\draw[directed] (10.5*\y-0.6*2.7*\y,-10*\y-0.6*1.5*\y) -- (10.5*\y,-10*\y);
\draw[directed] (9.5*\y,-8.7*\y) -- (9.5*\y-0.6*1.7*\y,-8.7*\y-0.6*2.8*\y);
\draw[directed] (11.5*\y,-11.1*\y) -- (10.5*\y,-10*\y);
\draw[directed][] (10.5*\y,-10*\y) -- (10.2*\y,-9*\y);
\draw[directed][] (10.2*\y,-9*\y) -- (9.5*\y,-8.7*\y);
\draw[directed] (9.5*\y,-8.7*\y) -- (8.5*\y,-7.6*\y);
\draw[directed] (10.2*\y,-9*\y) -- (11.2*\y,-8*\y);

\end{tikzpicture}
\end{center}

\caption{Singular part of the box integral written as the product of a vertex and a tree-level $5$-point diagram. The Jacobian $\frac{i}{8 \Delta_{BC}}$ is omitted in the picture. Note that, both here and in figure \ref{Example_of_on_shell_one_loop_box_diagram_how_to_perform_the_limit},
the faces of the dual diagram are mapped to the vertices of the Feynman diagram only after the former has been rotated by $90^{\circ}$ anticlockwise.}
\label{Box_diagram_after_the_cut}
\end{figure}

The subleading order in the integral expansion, $I_1$, does not present substantial difficulties in the computation. It is obtained by expanding the quantities linear in $(s-s_0)$ in the denominators of \eqref{Expression_for_I1_intitial_step_in_computing_the_box}, i.e.\ the terms containing $\tilde{l}^2$ and the second-order derivatives of the momenta. For example in the case of the $B'$-propagator we obtain
\begin{multline}
\frac{1}{\frac{d B'^2}{ds} + 2 B' \cdot \tilde{l}+\frac{1}{2} \frac{d^2 B'^2}{ds^2} (s-s_0) + (s-s_0) \tilde{l}^2+i \epsilon}\\
= \frac{1}{\frac{d B'^2}{ds} + 2 B' \cdot \tilde{l}+i \epsilon} - (s-s_0) \frac{\frac{1}{2} \frac{d^2 B'^2}{ds^2} + \tilde{l}^2}{\Bigl(\frac{d B'^2}{ds} + 2 B' \cdot \tilde{l}+i \epsilon\Bigr)^2} + \ldots
\end{multline}
where the second term on the RHS is the quantity we are interested in.
The part proportional to $\tilde{l}^2$ at the numerator gives a zero result after the integration. This is not surprising since from the change of variable \eqref{u_and_v_as_a_function_of_tilde_l} we see that $\tilde{l}^2$ needs to be a homogeneous polynomial of degree two in the $u$ and $v$ integration variables. Inverting \eqref{u_and_v_as_a_function_of_tilde_l} we find that $\tilde{l}^2 \sim 2 u v B \cdot C -u^2 m^2_C - v^2 m^2_B $. When we evaluate the integrand at the pole positions $u=-i \epsilon$, $v=-i \epsilon$ in the integration path,  we obtain $\tilde{l}^2 \sim \epsilon^2 \to 0$.
For this reason, we only consider the expansion coming from the second-order derivatives of the momenta that, expressed in terms of the $u$ and $v$ variables, is given by
\begin{equation}
\begin{split}
I_1&=-\frac{1}{(s-s_0)}\int  \frac{du\ dv}{(2 \pi)^2 8 i \Delta_{BC}} \frac{1}{u +i \epsilon} \ \frac{1}{v +i \epsilon} \\
&\times \frac{1}{\frac{d B'^2}{ds} -\frac{\Delta_{B'C}}{\Delta_{BC}} u  -\frac{\Delta_{B'B}}{\Delta_{BC}} v +i \epsilon} \ \frac{1}{\frac{d C'^2}{ds} -\frac{\Delta_{C'C}}{\Delta_{BC}} u  -\frac{\Delta_{C'B}}{\Delta_{BC}}v  +i \epsilon}\\
&\times \frac{1}{2} \Bigl[ \frac{d^2 B'^2}{ds^2} \frac{1}{\frac{d B'^2}{ds} -\frac{\Delta_{B'C}}{\Delta_{BC}} u  -\frac{\Delta_{B'B}}{\Delta_{BC}} v +i \epsilon} +  \frac{d^2 C'^2}{ds^2} \frac{1}{\frac{d C'^2}{ds} -\frac{\Delta_{C'C}}{\Delta_{BC}} u  -\frac{\Delta_{C'B}}{\Delta_{BC}}v  +i \epsilon} \Bigr].
\end{split}
\end{equation}
Once again the integration is  simple and can be performed by closing the $u$ and $v$ contours in the lower half complex plane
\begin{equation}
\label{subleading_term_to_the_box_integral_B_C_cut}
I_1=-\frac{1}{(s-s_0)}\frac{i}{ 8 \Delta_{BC}} \frac{1}{\frac{d B'^2}{ds}} \frac{1}{\frac{d C'^2}{ds} }\times \frac{1}{2} \biggl[ \frac{\frac{d^2 B'^2}{ds^2}}{\frac{d B'^2}{ds}}+ \frac{\frac{d^2 C'^2}{ds^2}}{\frac{d C'^2}{ds}}\biggr].
\end{equation}
Following the convention~\eqref{definition_of_singular_part}, by summing the leading and subleading terms \eqref{leading_term_to_the_box_integral_B_C_cut} and \eqref{subleading_term_to_the_box_integral_B_C_cut},
we see that the singular part of the box integral is 
\begin{multline}
\label{box_diagram_transformed_into_the_singular_part_of_a_tree_level_diagram_after_the_cuts}
I^{\text{(sing)}}=\frac{i}{ 8 \Delta_{BC}} \biggl[  \frac{1}{\frac{d B'^2}{ds}\Bigr|_{B,C} (s-s_0)+\frac{1}{2} \frac{d^2 B'^2}{ds^2}\Bigr|_{B,C} (s-s_0)^2}\\
\times \frac{1}{\frac{d C'^2}{ds}\Bigr|_{B,C} (s-s_0)+\frac{1}{2} \frac{d^2 C'^2}{ds^2}\Bigr|_{B,C} (s-s_0)^2}  \biggr]^{(\text{\tiny{sing}})}.
\end{multline}
In performing the integration we have broken the loop. The $B$- and $C$-propagators, having poles in the $u$- and $v$-variable inside the complex contour, have disappeared leaving only a flux term $\frac{i}{ 8 \Delta_{BC}}$ coming from the Jacobian in \eqref{jacobian_of_the_change_of_variables_box_integral_example}. The propagators with respect to which we took the residues have been cut and what remains is the product between a $3$-point vertex and a $5$-point diagram where the particles $B'$ and $C'$ propagate, as depicted in figure~\ref{Box_diagram_after_the_cut}. 
The Landau singularity corresponds exactly to the singular part of such a tree-level graph. The problem of computing the loop reduces then to a classical problem of evaluating the derivatives of the momenta around $s=s_0$ in a tree-level diagram. The correspondence is illustrated in figure~\ref{Box_diagram_after_the_cut}. 

We now show how to find the derivative of $B'^2$ with respect to $s$; the remaining computation can be performed similarly. We label by $\bar{U}_{bC}$ the angle between the sides $p_b$ and $C$ in the on-shell dual diagram in figure \ref{Example_of_on_shell_one_loop_box_diagram_how_to_perform_the_limit} and by $\bar{U}_{ab}$ the angle between the incoming momenta $p_a$ and $p_b$. Then we can write
\begin{equation}
\begin{split}
B'^2&=m^2_b+m^2_C-2m_b m_c \cos \bar{U}_{bC}\\
s&=m^2_a+m^2_b-2m_a m_b \cos \bar{U}_{ab}.
\end{split}
\end{equation}
Since all the sides of the triangle $\Delta_{aBC}$ (the small triangle on the top part of the parallelogram in figure \ref{Example_of_on_shell_one_loop_box_diagram_how_to_perform_the_limit}) are fixed we have
$
d \bar{U}_{ab}=d (\bar{U}_{aC} + \bar{U}_{bC}) = d \bar{U}_{bC},
$
therefore the derivative of $B'^2$ respect to $s$ at the pole is given by
\begin{equation}
\label{first_order_derivative_box_example_of_Bpr_respect_to_s}
\frac{dB'^2}{d s}= \frac{\frac{dB'^2}{d \bar{U}_{bC}}}{\frac{ds}{d \bar{U}_{ab}}} = \frac{2 i m_b m_c \sin \bar{U}_{bC}}{2 i m_a m_b \sin \bar{U}_{ab}}  = \frac{\Delta_{B' C}}{\Delta_{ab}}
\end{equation}
Similarly, it is possible to obtain
\begin{equation}
\label{second_and_first_order_derivatives_box_example}
\begin{split}
\frac{dC'^2}{d s}&= - \frac{\Delta_{B C'}}{\Delta_{ab}}\\
\frac{d^2B'^2}{d s^2}&=\frac{d^2C'^2}{d s^2} =\frac{p_b^2}{8}\frac{\Delta_{BC}}{\Delta^3_{ab}}.
\end{split}
\end{equation}

To get the Feynman diagram result we still need to multiply by the remaining $3$-point couplings, one for each vertex entering into the box. 
We split the Feynman diagram, as we did for the box integral, into a second-order singularity and a first-order one
$$
D^{\text{(sing)}}=D_2+D_1.
$$
Then using the area rule in \eqref{Connection_among_three_point_couplings_and_areas} and the expressions in \eqref{first_order_derivative_box_example_of_Bpr_respect_to_s}, \eqref{second_and_first_order_derivatives_box_example}, after having checked that the product of the $3$-point couplings is positive (this will be proved in the next section), we obtain
\begin{equation}
\label{D_written_in_terms_of_I_with_proper_three_point_couplings_everithing_written_explicitly}
\begin{split}
D_2&=-32 i \Bigl( \frac{\beta}{\sqrt{h}} \Bigr)^4 \frac{\Delta_{ab}^2\Delta_{B'C'}}{(s-s_0)^2}\\
D_1&=\Bigl( \frac{\beta}{\sqrt{h}} \Bigr)^4 \frac{i}{s-s_0} \times 2 \frac{\Delta_{BC}\Delta_{B'C'}}{\Delta_{B'C} \Delta_{BC'}} (\Delta_{BC'} - \Delta_{B' C}) p_b^2.
\end{split}
\end{equation}
The formulae in 
\eqref{D_written_in_terms_of_I_with_proper_three_point_couplings_everithing_written_explicitly} 
give the second- and first-order poles of the box diagram depicted in figure \ref{Example_of_on_shell_one_loop_box_diagram_how_to_perform_the_limit} at order $\beta^4$.
This result, however, is not universal; in particular, it involves the different triangle areas appearing in the on-shell diagram and therefore depends on the theory and process
under consideration. We are still far from a general formula similar to the bootstrapped result of \eqref{S_matrix_expansion_obtained_using_bootstrap_on_the_pole}. To get the full answer we need to sum over all the singular graphs of a bigger network \cite{Braden:1990wx} to which the box diagram so far studied belongs.

\subsection{The second-order pole network}
\begin{figure}
\hskip -10pt
\includegraphics[width=1.1\linewidth]{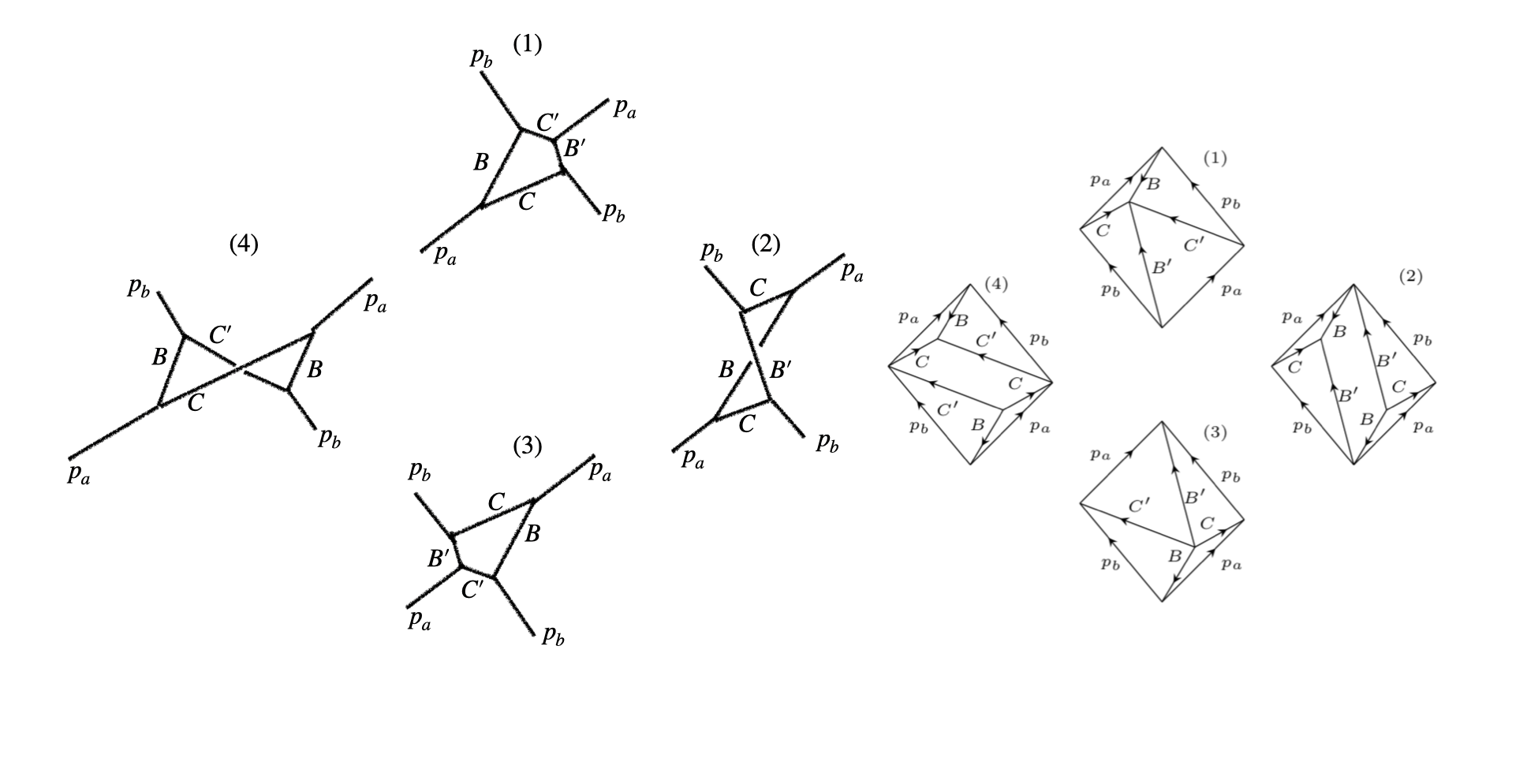}
 \caption{A network of singular Feynman diagrams (on the left) and their on-shell dual descriptions (on the right).}
\label{allowed_one_loop_box_network_on_shell_and_F_Diagram_description}
\end{figure}

Given one Feynman diagram contributing to the second-order pole, as is the case of the box graph previously studied, a network 
of four singular diagrams can be generated via the flipping rule. 
These diagrams are shown in figure~\ref{allowed_one_loop_box_network_on_shell_and_F_Diagram_description} and are  connected by changing internal propagators with type II flips. Suppose we start from diagram $(1)$
in figure~\ref{allowed_one_loop_box_network_on_shell_and_F_Diagram_description}, which is exactly the box diagram studied in the previous section. If we look at the RHS part of the on-shell description of the box, we note that it involves a $4$-point non-allowed process having for external states particles with momenta $B$, $B'$, $p_a$ and $p_b$, and for internal propagator $C'$. Since this sub-process contained in the loop is forbidden, we can apply a flip of type II, in which we cross the order of $B$ and $B'$ and we change the propagator from $C'$ to $C$. In this manner, we obtain a new singular diagram, $(2)$ in figure \ref{allowed_one_loop_box_network_on_shell_and_F_Diagram_description}. 
This is the same type of flip that connects the two tree-level diagrams in figure \ref{2_to_2_scattering_tree_level_simultaneous_poles_t_and_u_channels}.
This type of flipped move can be repeated until  we obtain the full network composed of four on-shell diagrams $(1)$, $(2)$, $(3)$ and $(4)$.
The reason why only flips of type II enter into the network, and not the other types discussed in \cite{Patrick_Davide_paper}, is that, as remarked earlier, at the second-order pole position no  on-shell bound states  propagate in the direct or crossed channels. Therefore 
in the parallelogram with sides equal to the external on-shell momenta, 
the two diagonals, which have lengths $\sqrt{s_0}$ and $\sqrt{2m_a^2+2m_b^2-s_0}$, do not correspond to any propagating on-shell bound states. This holds for all rapidity values corresponding to even order poles. 

Each diagram can be computed in a similar way to that  explained in the previous section. Given a generic diagram, we only need to understand what are the internal momenta with respect to which all the others are negative linear combinations, in such a way to have a complex contour enclosing nicely the poles in the $u$ and $v$ integration variables. As we have shown in \eqref{Bprime_and_Cprime_as_linear_combinations_of_B_and_C} such momenta are $B$ and $C$ in  diagram $(1)$, and  also in diagram $(3)$ since it is a rotation of $(1)$. This means that the two diagrams are equivalent and both are obtained by cutting the $B$ and $C$ propagators, and inserting a flux factor coming from the Jacobian of the transformation of variables. They return, up to extra three-point vertices, the expression in \eqref{box_diagram_transformed_into_the_singular_part_of_a_tree_level_diagram_after_the_cuts}, which is proportional to a singular tree-level diagram with five external legs.
Similarly in  diagram $(2)$ the momenta with respect to which all the remaining vectors are negative linear combinations can be read from its on-shell description on the RHS of figure \ref{allowed_one_loop_box_network_on_shell_and_F_Diagram_description}. They are $B$ and $B'$ with respect to which $C$ is given by
$$
C=-\frac{\Delta_{B'C}}{\Delta_{BB'}} B - \frac{\Delta_{BC}}{\Delta_{BB'}} B'.
$$
This means that the $B$ and $B'$ propagators are cut once we perform the loop integral and the singular part of the crossed box diagram $(2)$ is broken into the product of two equal $4$-point processes having as external on-shell momenta $p_a$, $p_b$, $B$ and $B'$. It is given by 
\begin{multline}
\label{diagram_I2_final_result_non_explicit_singular_part}
I^{(2)}= \frac{i}{8 \Delta_{BB'}} \biggl[  \frac{1}{\frac{d C^2}{ds}\Bigr|_{B,B'} (s-s_0)+\frac{1}{2} \frac{d^2 C^2}{ds^2}\Bigr|_{B,B'} (s-s_0)^2} \\
\times \frac{1}{\frac{d C^2}{ds}\Bigr|_{B,B'} (s-s_0)+\frac{1}{2} \frac{d^2 C^2}{ds^2}\Bigr|_{B,B'} (s-s_0)^2}  \biggr]^{(\text{\tiny{sing}})}.
\end{multline}
Similarly, the singular part of the integral associated with the fourth diagram is
\begin{multline}
\label{diagram_I4_final_result_non_explicit_singular_part}
I^{(4)}= \frac{i}{8 \Delta_{CC'}} \biggl[  \frac{1}{\frac{d B^2}{ds}\Bigr|_{C,C'} (s-s_0)+\frac{1}{2} \frac{d^2 B^2}{ds^2}\Bigr|_{C,C'} (s-s_0)^2} \\
\times \frac{1}{\frac{d B^2}{ds}\Bigr|_{C,C'} (s-s_0)+\frac{1}{2} \frac{d^2 B^2}{ds^2}\Bigr|_{C,C'} (s-s_0)^2}  \biggr]^{(\text{\tiny{sing}})}.
\end{multline}
Now we have to pay attention to one thing. When expand these last two expressions in $(s-s_0)$ we need to take into account that the areas $\Delta_{BB'}$ and $\Delta_{CC'}$, coming from the Jacobian of the change of variables, depend on $s$. 
Indeed the sides composing the triangles $\Delta_{BB'}$ and $\Delta_{CC'}$ are respectively $m_B$, $m_{B'}$, $\sqrt{s}$ and $m_C$, $m_{C'}$, $\sqrt{2m_a^2+2m_b^2-s}$. Since we want to evaluate the $s$-expansion of the integrals we also need to Taylor expand  such Jacobian factors. In the case of $\Delta_{BB'}(s)$ we note that
$$
\Delta_{BB'}(s)=\frac{1}{2}m_B m_{B'} \sin \bar{U}_{BB'}(s)
$$
where $\bar{U}_{BB'}(s)$ depends on $s$ through
$$
s=m_B^2+m_{B'}^2 - 2 m_B m_{B'} \cos \bar{U}_{BB'}(s),
$$
where we follow the same convention previously used to define the angle. Therefore we obtain
$$
\frac{d\Delta_{BB'}(s)}{ds}=\frac{\frac{d\Delta_{BB'}}{d \bar{U}_{B B'}}}{\frac{d s}{d \bar{U}_{B B'}}}=\frac{1}{4} \operatorname{cotan} \bar{U}_{BB'}(s)
$$
and the expansion of the flux factor in front of \eqref{diagram_I2_final_result_non_explicit_singular_part} is  given by
$$
\frac{1}{\Delta_{BB'}(s)}= \frac{1}{\Delta_{BB'}(s_0)} \Bigl[ 1 -\frac{(s-s_0)}{4 \Delta_{BB'}(s_0)} \operatorname{cotan} \bar{U}_{BB'}(s_0) +\ldots \Bigr].
$$
A similar relation can be found also for $\Delta_{CC'}(s)$. Expanding the expressions in \eqref{diagram_I2_final_result_non_explicit_singular_part}, \eqref{diagram_I4_final_result_non_explicit_singular_part} with these further considerations, inserting additional $3$-point vertices as given in \eqref{Connection_among_three_point_couplings_and_areas} and substituting the correct values of the momentum derivatives we obtain that the contributions to the double and single poles in $(s-s_0)$, given by the diagram in figure~\ref{allowed_one_loop_box_network_on_shell_and_F_Diagram_description}, are 
\begin{equation}
\label{second_order_contributions_s_expansion_D_forms}
\begin{split}
&D_2^{(1)}=D_2^{(3)}=-32 i \Bigl( \frac{\beta}{\sqrt{h}} \Bigr)^4 \frac{\Delta_{ab}^2\Delta_{B'C'}}{(s-s_0)^2}\\
&D_2^{(2)} = 32 i \Bigl( \frac{\beta}{\sqrt{h}} \Bigr)^4 \frac{\Delta_{ab}^2\Delta_{BB'}}{(s-s_0)^2}\\
&D_2^{(4)} = 32 i \Bigl( \frac{\beta}{\sqrt{h}} \Bigr)^4 \frac{\Delta_{ab}^2\Delta_{CC'}}{(s-s_0)^2}.
\end{split}
\end{equation}
and
\begin{equation}
\label{first_order_contributions_s_expansion_D_forms}
\begin{split}
D_1^{(1)}=D_1^{(3)} &= \Bigl( \frac{\beta}{\sqrt{h}} \Bigr)^4 \frac{i}{s-s_0} \times 2 \frac{\Delta_{BC}\Delta_{B'C'}}{\Delta_{B'C} \Delta_{BC'}} (\Delta_{BC'} - \Delta_{B' C}) p_b^2\\
D_1^{(2)}&=\Bigl( \frac{\beta}{\sqrt{h}} \Bigr)^4 \frac{i}{s-s_0} \times \biggl[-4 \frac{\Delta_{B'C}\Delta_{BC'}}{\Delta_{BC}} p_a^2 -4 \frac{\Delta_{BC}\Delta_{B'C'}}{\Delta_{B'C}} p_b^2 -4 \Delta_{ab} p_a \cdot p_b\biggr]\\
D_1^{(4)}&=\Bigl( \frac{\beta}{\sqrt{h}} \Bigr)^4 \frac{i}{s-s_0} \times \biggl[4 \frac{\Delta_{B'C}\Delta_{BC'}}{\Delta_{BC}} p_a^2 +4 \frac{\Delta_{BC}\Delta_{B'C'}}{\Delta_{BC'}} p_b^2 -4 \Delta_{ab} p_a \cdot p_b\biggr].
\end{split}
\end{equation}
In these two sets of expressions  the subscript index ($2$ or $1$) indicates the order of the pole in $(s-s_0)$,  a superscript index ($(1)$,$(2)$,$(3)$ or $(4)$) labels the  contributing diagram, and all triangle areas $\Delta$ are evaluated at $s=s_0$.

A necessary ingredient to reproduce the correct sign in each term in \eqref{second_order_contributions_s_expansion_D_forms} and \eqref{first_order_contributions_s_expansion_D_forms} is to know the product between the different $3$-point couplings.
To understand, for each Feynman diagram, what sign results from multiplying the different couplings together we note that the graphs in figure~\ref{allowed_one_loop_box_network_on_shell_and_F_Diagram_description} are connected by type II flips. This implies that the product between the $f$-functions, defined in \eqref{Connection_among_three_point_couplings_and_areas},
does not change passing from one diagram to one other:
\begin{equation}
\label{products_between_f_functions_entering_the_2nd_order_network}
\begin{split}
f_{Ba\bar{C}} f_{Cb\bar{B}'} f_{B' \bar{a} \bar{C}'} f_{C' \bar{b} \bar{B}}&=|f_{Ba\bar{C}}|^2 |f_{Cb\bar{B}'}|^2=f_{\bar{B} \bar{a}C} f_{\bar{C}\bar{b}B'} f_{\bar{B}' a C'  } f_{\bar{C}' b B}\\
&=|f_{\bar{B} \bar{a}C}|^2 |f_{\bar{C}' b B}|^2=\Bigl(\frac{4\beta}{\sqrt{h}} \Bigr)^4.
\end{split}
\end{equation}
Since the common value is certainly positive in diagrams $(2)$ and $(4)$, we conclude that the product of the different $f$-functions, entering in the $3$-point couplings, is positive and satisfies~\eqref{products_between_f_functions_entering_the_2nd_order_network}.

The sum of the double pole contributions in \eqref{second_order_contributions_s_expansion_D_forms}  simplifies once we note  that $\Delta_{B'C'}-\Delta_{BB'}=\Delta_{BC'}$, $\Delta_{B'C'}-\Delta_{CC'}=\Delta_{B'C}$ and $\Delta_{BC'}+\Delta_{B'C}=\Delta_{ab}$.
Summing the four contributions in \eqref{second_order_contributions_s_expansion_D_forms} we obtain
\begin{equation}
\label{sum_of_second_order_contributions_in_s}
\sum_{n=1}^4 D_2^{(n)}=-32 i \Bigl( \frac{\beta}{\sqrt{h}} \Bigr)^4 \frac{\Delta_{ab}^3}{(s-s_0)^2}.
\end{equation}
Similarly the sum of the quantities in \eqref{first_order_contributions_s_expansion_D_forms} returns a simple expression given by
\begin{equation}
\label{sum_of_first_order_contributions_in_s_coming_from_double_pole_diagrams}
\sum_{n=1}^4 D_1^{(n)}= -8 i  \Bigl( \frac{\beta}{\sqrt{h}} \Bigr)^4 \frac{\Delta_{ab}}{(s-s_0)} p_a \cdot p_b.
\end{equation}

It is worth noting that in contrast to the tree level situation, where singularities in sums of Feynman diagrams connected by one flip cancel in two to two non-diagonal processes, here the sums of the poles appearing in the different contributions in~\eqref{second_order_contributions_s_expansion_D_forms} do not sum to zero. The reason why this happens is that when we cut loop diagrams into products of tree-level graphs, the loops are broken in different ways from one diagram to one other. For example,  diagram $(2)$ in figure~\ref{allowed_one_loop_box_network_on_shell_and_F_Diagram_description} is cut into two non-diagonal 2 to 2 tree-level graphs, having removed the propagators $B$ and $B'$ thorough the cut. However, diagram $(1)$ is cut into one $3$-point vertex and a $5$-point tree-level diagram. Therefore the values of diagrams $(1)$ and $(2)$ on the pole do not differ by two different on-shell $4$-point Feynman diagrams of the form in \ref{2_to_2_scattering_tree_level_simultaneous_poles_t_and_u_channels}, since in diagram $(1)$ the tree-level diagram appearing after the cut contains $5$ external on-shell particles. We will investigate how to recover possible simplifications in the evaluation of loop diagrams connected by flips, similarly to what we observe at the tree level, in a companion paper~\cite{Second_loop_paper_sagex}. 

The expressions in~\eqref{sum_of_second_order_contributions_in_s} and~\eqref{sum_of_first_order_contributions_in_s_coming_from_double_pole_diagrams} arise by summing just one network of singular diagrams. 
Even though we do not have a universal proof of this fact, it turns out that if the highest order pole in~\eqref{Expansion_of_S_around_an_arbitrary_2N_order_pole} is $2N$ there are a total of $N$ different (and disconnected) networks of the type depicted in figure~\ref{allowed_one_loop_box_network_on_shell_and_F_Diagram_description} that are on-shell at the pole location\footnote{Related observations about the counting of subtilings can be found in \cite{Braden:1990wx}.}. Each network contributes to the double and single poles with factors~\eqref{sum_of_second_order_contributions_in_s} and~\eqref{sum_of_first_order_contributions_in_s_coming_from_double_pole_diagrams} respectively so that, for a pole of highest order $P=2N$, we obtain
\begin{equation}
\label{expansion_of_N_copies_of_the_network}
   N\Bigl[ -32 i \Bigl( \frac{\beta}{\sqrt{h}} \Bigr)^4 \frac{\Delta_{ab}^3}{(s-s_0)^2} -8 i  \Bigl( \frac{\beta}{\sqrt{h}} \Bigr)^4 \frac{\Delta_{ab}}{(s-s_0)} p_a \cdot p_b \Bigr].
\end{equation}
Before writing~\eqref{expansion_of_N_copies_of_the_network} in terms of the rapidity to compare it
with the Laurent expansion~\eqref{Expansion_of_S_around_an_arbitrary_2N_order_pole}, we  need to consider all the remaining Feynman diagrams yielding simple poles in $(s-s_0)$ at one loop.

\subsection{The first-order pole network}

In the previous section we have evaluated, in \eqref{expansion_of_N_copies_of_the_network}, the sum of Feynman diagrams contributing to the pole at order $(s-s_0)^{-2}$ , and their subleading expansion of order $(s-s_0)^{-1}$.
In order to reproduce the correct value of $b_1$ in~\eqref{Expansion_of_S_around_an_arbitrary_2N_order_pole}, we still need to add to~\eqref{expansion_of_N_copies_of_the_network} all the loop diagrams having simple singularities at $s=s_0$, i.e.\  diagrams which have only three internal on-shell propagators when $s=s_0$. 
We remark that the possibility of having order one singularities generated by on-shell bound states in the direct or crossed channel is excluded since, as already remarked in the introduction, the coefficient $a_1$ in~\eqref{Expansion_of_S_around_an_arbitrary_2N_order_pole} is null, meaning that the tree-level S-matrix is non-singular at the pole rapidity value. Due to this fact, the only possible graphs contributing to the first order pole in the Laurent expansion of the S-matrix are anomalous threshold diagrams: they are the configurations where we can find a point `${\bf o}$' internally to the parallelogram defined by the on-shell momenta $p_a$, $p_b$,  such that three of the four segments obtained by connecting  `${\bf o}$' with the vertices of the parallelogram have lengths equal to the masses of the propagating particles.
Graphs of this type are all  candidates to yield a  pole of order one. However this is a necessary, but not sufficient, condition for a diagram to contribute to the singularity. At this point, we need to exclude all the diagrams that, due to their geometry, have zero residue at $s=s_0$. This is the case if the graph is of the type depicted in the first row of figure \ref{Example_of_null_diagram_contrinuting_to_the_simple_pole}. As explained in section \ref{subsection_explanation_of_the_method}, since the loop integration variable carries two degrees of freedom, we can choose two among the three propagators which diverge at $s=s_0$, and have them on-shell at $l=(0,0)$
also away from the pole position. Referring to figure \ref{Example_of_null_diagram_contrinuting_to_the_simple_pole} we can set $F^2=m^2_F$ and $G^2=m^2_G$. Since all the freedom has been already used to fix such momenta, now $E$ is completely determined; it is not on-shell for $s$ arbitrary but it reaches its mass-shell value only on the pole
$$
E^2-m^2_E=\frac{dE^2}{ds}\Bigr|_{F,G}(s-s_0).
$$
After rescaling $l=(s-s_0) \tilde{l}$ and adopting the usual change of variables
$$
2 F \cdot \tilde{l} = u  \ \ \ , \ \ \ 2 G \cdot \tilde{l} = v \ \ \ , \ \ \ 2 E \cdot \tilde{l} = \frac{\Delta_{EG}}{\Delta_{FG}} u + \frac{\Delta_{EF}}{\Delta_{FG}} v,
$$
the residue at the pole is  given by
$$
\frac{1}{8i \Delta_{FG}}\int \frac{du dv}{(2 \pi)^2} \frac{1}{u+i\epsilon} \  \frac{1}{v+i\epsilon} \ \frac{1}{\frac{dE^2}{ds}+\frac{\Delta_{EG}}{\Delta_{FG}} u +\frac{\Delta_{EF}}{\Delta_{FG}} v  +i\epsilon} \times FD_{abGF}(s_0),
$$
where we have labelled with $FD_{abGF}(s_0)$ the part of the loop which is finite at the pole, whatever combination of vertices and propagators it corresponds to. 
\begin{figure}
\medskip
\begin{center}

\begin{tikzpicture}
\tikzmath{\x = 1;\y = 0.7;}

%Diagram that does not contribute
\draw[directed] (0.8*\y,1*\y-5*\y) -- (2.3*\y,2.5*\y-5*\y);
\draw[directed] (2.3*\y,-0.8*\y-5*\y) -- (0.8*\y,1*\y-5*\y);
\draw[directed] (2.3*\y,-0.8*\y-5*\y) -- (3.8*\y,0.7*\y-5*\y);
\draw[directed] (3.8*\y,0.7*\y-5*\y) -- (2.3*\y,2.5*\y-5*\y);
\draw[directed] (0.8*\y,1*\y-5*\y) -- (2.5*\y,1.2*\y-5*\y);
\draw[directed] (2.3*\y,-0.8*\y-5*\y) -- (2.5*\y,1.2*\y-5*\y);
\draw[directed] (2.3*\y,2.5*\y-5*\y) -- (2.5*\y,1.2*\y-5*\y);

\filldraw[black] (0.8*\y,2*\y-5*\y)  node[anchor=west] {\tiny{$p_a$}};
\filldraw[black] (1*\y,0*\y-5*\y)  node[anchor=west] {\tiny{$p_b$}};
\filldraw[black] (3.2*\y,0*\y-5*\y)  node[anchor=west] {\tiny{$p_a$}};
\filldraw[black] (3.2*\y,1.6*\y-5*\y)  node[anchor=west] {\tiny{$p_b$}};
\filldraw[black] (1.9*\y,1.8*\y-5*\y)  node[anchor=west] {\tiny{$F$}};
\filldraw[black] (1.5*\y,1.4*\y-5*\y)  node[anchor=west] {\tiny{$E$}};
\filldraw[black] (1.9*\y,0.6*\y-5*\y)  node[anchor=west] {\tiny{$G$}};
\filldraw[black] (2.35*\y,1.2*\y-5*\y)  node[anchor=west] {\tiny{${\bf o}$}};

\filldraw[black] (4.3*\y,-4*\y)  node[anchor=west] {\small{$= \ \frac{1}{(s-s_0)} \frac{1}{8i \Delta_{FG}}\int \frac{du dv}{(2 \pi)^2} \frac{1}{u+i\epsilon} \  \frac{1}{v+i\epsilon} \ \frac{1}{\frac{dE^2}{ds}+\frac{\Delta_{EG}}{\Delta_{FG}} u +\frac{\Delta_{EF}}{\Delta_{FG}} v  +i\epsilon} \times FD_{abGF}(s_0)$}};

%Diagram that contributes
\draw[directed] (0.8*\y,1*\y-10*\y) -- (2.3*\y,2.5*\y-10*\y);
\draw[directed] (0.8*\y,1*\y-10*\y) -- (1.7*\y,1.5*\y-10*\y);
\draw[directed] (2.3*\y,2.5*\y-10*\y) -- (1.7*\y,1.5*\y-10*\y);
\draw[directed] (2.3*\y,-0.8*\y-10*\y) -- (0.8*\y,1*\y-10*\y);
\draw[directed] (2.3*\y,-0.8*\y-10*\y) -- (3.8*\y,0.7*\y-10*\y);
\draw[directed] (3.8*\y,0.7*\y-10*\y) -- (2.3*\y,2.5*\y-10*\y);
\draw[directed] (2.3*\y,-0.8*\y-10*\y) -- (1.7*\y,1.5*\y-10*\y);

\filldraw[black] (0.8*\y,2*\y-10*\y)  node[anchor=west] {\tiny{$p_a$}};
\filldraw[black] (1*\y,0*\y-10*\y)  node[anchor=west] {\tiny{$p_b$}};
\filldraw[black] (3.2*\y,0*\y-10*\y)  node[anchor=west] {\tiny{$p_a$}};
\filldraw[black] (3.2*\y,1.6*\y-10*\y)  node[anchor=west] {\tiny{$p_b$}};
\filldraw[black] (1.9*\y,1.8*\y-10*\y)  node[anchor=west] {\tiny{$F$}};
\filldraw[black] (1*\y,1.1*\y-10*\y)  node[anchor=west] {\tiny{$E$}};
\filldraw[black] (1.9*\y,0.6*\y-10*\y)  node[anchor=west] {\tiny{$G$}};
\filldraw[black] (1.6*\y,1.4*\y-10*\y)  node[anchor=west] {\tiny{${\bf o}$}};

\filldraw[black] (4.3*\y,-9*\y)  node[anchor=west] {\small{$= \ \frac{1}{(s-s_0)} \frac{1}{8i \Delta_{FG}}\int \frac{du dv}{(2 \pi)^2} \frac{1}{u+i\epsilon} \  \frac{1}{v+i\epsilon} \ \frac{1}{\frac{dE^2}{ds}-\frac{\Delta_{EG}}{\Delta_{FG}} u -\frac{\Delta_{EF}}{\Delta_{FG}} v  +i\epsilon} \times FD_{abGF}(s_0)$}};
\end{tikzpicture}
\end{center}
\caption{On the first row is an example of a diagram that could in principle contribute to the pole at  order $(s-s_0)^{-1}$ but it is zero. This is because all the simple poles in the $u$- and $v$-variable lie in the same half-plane, since $E$ is a positive linear combination of $F$ and $G$. The diagram on the second row instead contributes to the pole with a result different from zero. In both the expressions, $FD_{abGF}(s_0)$ represents any tree level diagram with external insertions $p_a$, $p_b$, $G$ and $F$ which is finite at the position $s=s_0$.  }
\label{Example_of_null_diagram_contrinuting_to_the_simple_pole}
\end{figure}
We see that in the expression above there exists a $(u,v)$-path in the complex plane which does not contain any poles and the integral is trivially equal to zero. This is because the vector $E$ is a positive linear combination of $F$ and $G$. To find the Feynman diagrams which do contribute to the pole at  order $(s-s_0)^{-1}$ with non-zero integrals the following condition needs to be satisfied: if $A_1$, $A_2$ and $A_3$ are the internal momenta becoming on-shell at the pole they need to satisfy 
\begin{equation}
\label{condition_to_havo_non_zero_residues_in_pole_of_order_one_diagrams}
\sum_{i=1}^3 \gamma_i A_i=0 \ \ \ \text{with} \ \ \ \gamma_i > 0 \ \forall \ i=1,2,3.
\end{equation}
In other words, any momentum can be expressed as a negative linear combination of the other two.
This condition is satisfied only if the angles defined between the arrowheads pointing towards `${\bf o}$' are all less than or equal to $180^\circ$, as shown in the diagram in the second row of figure~\ref{Example_of_null_diagram_contrinuting_to_the_simple_pole}. In this situation, we see that $E$ is indeed a negative linear combination of $F$ and $G$.
Therefore, the problem of finding graphs contributing to the simple pole amounts to finding all the Feynman diagrams whose internal on-shell momenta respect the condition \eqref{condition_to_havo_non_zero_residues_in_pole_of_order_one_diagrams}.

For a pole of maximal order $2N$ in~\eqref{Expansion_of_S_around_an_arbitrary_2N_order_pole}, we can organise the Feynman diagrams contributing with anomalous simple poles at the order $\beta^4$ into $N$ separate sets, one for each network of the type in figure~\ref{allowed_one_loop_box_network_on_shell_and_F_Diagram_description}. Let us suppose that the graph in the second row in figure~\ref{Example_of_null_diagram_contrinuting_to_the_simple_pole} contributes to the pole at order $(s-s_0)^{-1}$, having a point in the integration region where the internal particles $\{E,F,G\}$ are on-shell. In such a situation it has to be possible to tile the RHS part of the diagram, as we did for the LHS, with the same particles $\{E,F,G\}$, generating a graph of the same type of diagram $(2)$ in the network~\ref{allowed_one_loop_box_network_on_shell_and_F_Diagram_description}. Therefore the existence of the diagram in the second row of figure~\ref{Example_of_null_diagram_contrinuting_to_the_simple_pole} implies the existence of a diagram of the same form of diagram $(2)$ in 
one of the $N$ disjoint networks of type in~\ref{allowed_one_loop_box_network_on_shell_and_F_Diagram_description}. Since all the $N$ networks have the same structure it is not restrictive to assume that $\{E,F,G\}$ belong to the network in figure~\ref{allowed_one_loop_box_network_on_shell_and_F_Diagram_description}, and therefore are $\{C,B,B'\}$. The same argument can be repeated identically for the situation in which the on-shell internal particles are $\{C, C', B\}$. 

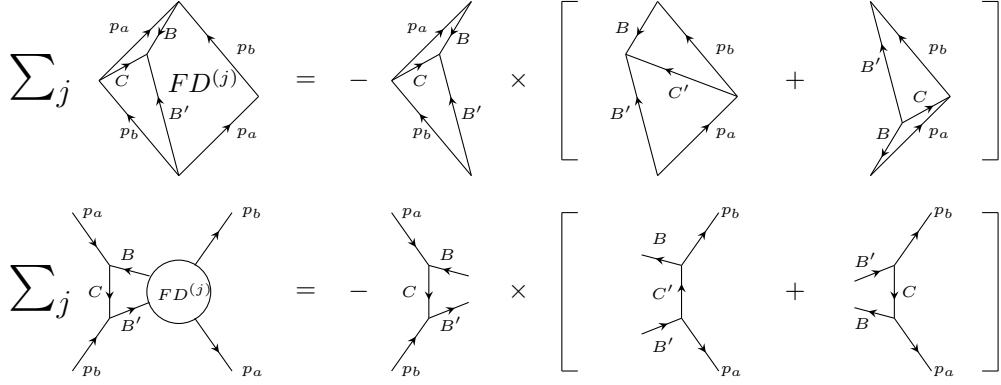
\begin{figure}
\medskip
\begin{center}
\begin{tikzpicture}
\tikzmath{\x = 1;\y = 0.7;}

\filldraw[black] (-1.1*\y,-4*\y)  node[anchor=west] {\LARGE{$\sum_j$}};

%On-shell box diagram               
\draw[directed] (0.8*\y,1*\y-5*\y) -- (2.3*\y,2.5*\y-5*\y);
\draw[directed] (0.8*\y,1*\y-5*\y) -- (1.7*\y,1.5*\y-5*\y);
\draw[directed] (2.3*\y,2.5*\y-5*\y) -- (1.7*\y,1.5*\y-5*\y);
\draw[directed] (2.3*\y,-0.8*\y-5*\y) -- (0.8*\y,1*\y-5*\y);
\draw[directed] (2.3*\y,-0.8*\y-5*\y) -- (1.7*\y,1.5*\y-5*\y);
\draw[directed] (2.3*\y,-0.8*\y-5*\y) -- (3.8*\y,0.7*\y-5*\y);
\draw[directed] (3.8*\y,0.7*\y-5*\y) -- (2.3*\y,2.5*\y-5*\y);

\filldraw[black] (0.8*\y,2*\y-5*\y)  node[anchor=west] {\tiny{$p_a$}};
\filldraw[black] (1*\y,0*\y-5*\y)  node[anchor=west] {\tiny{$p_b$}};
\filldraw[black] (3.2*\y,0*\y-5*\y)  node[anchor=west] {\tiny{$p_a$}};
\filldraw[black] (3.2*\y,1.6*\y-5*\y)  node[anchor=west] {\tiny{$p_b$}};
\filldraw[black] (0.9*\y,1*\y-5*\y)  node[anchor=west] {\tiny{$C$}};
\filldraw[black] (1.9*\y,0.4*\y-5*\y)  node[anchor=west] {\tiny{$B'$}};
\filldraw[black] (1.8*\y,1.9*\y-5*\y)  node[anchor=west] {\tiny{$B$}};

\filldraw[black] (1.9*\y,-4*\y)  node[anchor=west] {\small{$FD^{(j)}$}};

\filldraw[black] (4.3*\y,-4*\y)  node[anchor=west] {\small{$=$}};

%Box diagram               
\filldraw[black] (-1.1*\y,-8*\y)  node[anchor=west] {\LARGE{$\sum_j$}};

\draw[directed] (0.3*\y,-6.5*\y) -- (1*\y,-7.5*\y);
\draw[directed] (1*\y,-7.5*\y) -- (1*\y,-8.5*\y);
\draw[directed] (0.3*\y,-9.5*\y) -- (1*\y,-8.5*\y);
\draw (2.3*\y,-8*\y) circle (0.6*\y);
\draw[directed] (1*\y,-8.5*\y) -- (1.75*\y,-8.2*\y);
\draw[directed] (1.75*\y,-7.7*\y) -- (1*\y,-7.5*\y);
\draw[directed] (2.6*\y,-8.5*\y) -- (3.3*\y,-9.5*\y);
\draw[directed] (2.6*\y,-7.5*\y) -- (3.3*\y,-6.5*\y);

\filldraw[black] (0.3*\y,-6.5*\y)  node[anchor=west] {\tiny{$p_a$}};
\filldraw[black] (0.4*\y,-8*\y)  node[anchor=west] {\tiny{$C$}};
\filldraw[black] (0.3*\y,-9.5*\y)  node[anchor=west] {\tiny{$p_b$}};
\filldraw[black] (1*\y,-7.3*\y)  node[anchor=west] {\tiny{$B$}};
\filldraw[black] (1*\y,-8.6*\y)  node[anchor=west] {\tiny{$B'$}};
\filldraw[black] (1.7*\y,-8*\y)  node[anchor=west] {\tiny{$FD^{(j)}$}};
\filldraw[black] (3.3*\y,-6.5*\y)  node[anchor=west] {\tiny{$p_b$}};
\filldraw[black] (3.3*\y,-9.5*\y)  node[anchor=west] {\tiny{$p_a$}};

\filldraw[black] (4.3*\y,-8*\y)  node[anchor=west] {\small{$=$}};

%Cut On-shell box diagram term 1           
\filldraw[black] (5.3*\y,-4*\y)  node[anchor=west] {\small{$-$}};

\draw[directed] (0.3*\y+6*\y,1*\y-5*\y) -- (1.8*\y+6*\y,2.5*\y-5*\y);
\draw[directed] (0.3*\y+6*\y,1*\y-5*\y) -- (1.2*\y+6*\y,1.5*\y-5*\y);
\draw[directed] (1.8*\y+6*\y,2.5*\y-5*\y) -- (1.2*\y+6*\y,1.5*\y-5*\y);
\draw[directed] (1.8*\y+6*\y,-0.8*\y-5*\y) -- (0.3*\y+6*\y,1*\y-5*\y);
\draw[directed] (1.8*\y+6*\y,-0.8*\y-5*\y) -- (1.2*\y+6*\y,1.5*\y-5*\y);

\filldraw[black] (0.6*\y+5.8*\y,2*\y-5*\y)  node[anchor=west] {\tiny{$p_a$}};
\filldraw[black] (1*\y+5.5*\y,0*\y-5*\y)  node[anchor=west] {\tiny{$p_b$}};
\filldraw[black] (1.9*\y+5.5*\y,0.4*\y-5*\y)  node[anchor=west] {\tiny{$B'$}};
\filldraw[black] (1.8*\y+5.5*\y,1.8*\y-5*\y)  node[anchor=west] {\tiny{$B$}};
\filldraw[black] (1*\y+5.5*\y,1*\y-5*\y)  node[anchor=west] {\tiny{$C$}};

\filldraw[black] (8.3*\y,-4*\y)  node[anchor=west] {\small{$\times$}};

\draw[] (9.5*\y,-5.5*\y) -- (9.5*\y,-2.5*\y);
\draw[] (9.5*\y,-5.5*\y) -- (9.8*\y,-5.5*\y);
\draw[] (9.5*\y,-2.5*\y) -- (9.8*\y,-2.5*\y);

\draw[directed] (2.3*\y+9*\y,2.5*\y-5*\y) -- (1.7*\y+9*\y,1.5*\y-5*\y);
\draw[directed] (2.3*\y+9*\y,-0.8*\y-5*\y) -- (1.7*\y+9*\y,1.5*\y-5*\y);
\draw[directed] (2.3*\y+9*\y,-0.8*\y-5*\y) -- (3.8*\y+9*\y,0.7*\y-5*\y);
\draw[directed] (3.8*\y+9*\y,0.7*\y-5*\y) -- (2.3*\y+9*\y,2.5*\y-5*\y);
\draw[directed] (3.8*\y+9*\y, 0.7*\y-5*\y) -- (1.7*\y+9*\y,1.5*\y-5*\y);

\filldraw[black] (3.2*\y+9*\y,0*\y-5*\y)  node[anchor=west] {\tiny{$p_a$}};
\filldraw[black] (3.2*\y+9*\y,1.6*\y-5*\y)  node[anchor=west] {\tiny{$p_b$}};
\filldraw[black] (1.9*\y+8.3*\y,0.4*\y-5*\y)  node[anchor=west] {\tiny{$B'$}};
\filldraw[black] (2*\y+8.3*\y,2*\y-5*\y)  node[anchor=west] {\tiny{$B$}};
\filldraw[black] (3*\y+8.3*\y,0.8*\y-5*\y)  node[anchor=west] {\tiny{$C'$}};

%Cut  box diagram term 1      
\filldraw[black] (5.3*\y,-8*\y)  node[anchor=west] {\small{$-$}};

\draw[directed] (0.3*\y+6*\y,-6.5*\y) -- (1*\y+6*\y,-7.5*\y);
\draw[directed] (1*\y+6*\y,-7.5*\y) -- (1*\y+6*\y,-8.5*\y);
\draw[directed] (0.3*\y+6*\y,-9.5*\y) -- (1*\y+6*\y,-8.5*\y);
\draw[directed] (1*\y+6*\y,-8.5*\y) -- (1.75*\y+6*\y,-8.2*\y);
\draw[directed] (1.75*\y+6*\y,-7.7*\y) -- (1*\y+6*\y,-7.5*\y);

\filldraw[black] (0.3*\y+6*\y,-6.5*\y)  node[anchor=west] {\tiny{$p_a$}};
\filldraw[black] (0.3*\y+6*\y,-9.5*\y)  node[anchor=west] {\tiny{$p_b$}};
\filldraw[black] (1*\y+6*\y,-7.3*\y)  node[anchor=west] {\tiny{$B$}};
\filldraw[black] (0.3*\y+6*\y,-8*\y)  node[anchor=west] {\tiny{$C$}};
\filldraw[black] (1*\y+6*\y,-8.6*\y)  node[anchor=west] {\tiny{$B'$}};

\filldraw[black] (8.3*\y,-8*\y)  node[anchor=west] {\small{$\times$}};
\draw[] (9.5*\y,-9.5*\y) -- (9.5*\y,-6.5*\y);
\draw[] (9.5*\y,-9.5*\y) -- (9.8*\y,-9.5*\y);
\draw[] (9.5*\y,-6.5*\y) -- (9.8*\y,-6.5*\y);

\draw[directed] (1*\y+10*\y,-8.8*\y) -- (1.75*\y+10*\y,-8.5*\y);
\draw[directed] (1.75*\y+10*\y,-7.5*\y) -- (1*\y+10*\y,-7.3*\y);
\draw[directed] (1.75*\y+10*\y,-8.5*\y) -- (1.75*\y+10*\y,-7.5*\y);
\draw[directed] (1.75*\y+10*\y,-8.5*\y) -- (2.45*\y+10*\y,-9.5*\y);
\draw[directed] (1.75*\y+10*\y,-7.5*\y) -- (2.45*\y+10*\y,-6.5*\y);

\filldraw[black] (1*\y+10*\y,-7*\y)  node[anchor=west] {\tiny{$B$}};
\filldraw[black] (1*\y+10*\y,-9*\y)  node[anchor=west] {\tiny{$B'$}};
\filldraw[black] (1*\y+10*\y,-8*\y)  node[anchor=west] {\tiny{$C'$}};
\filldraw[black] (2.3*\y+10*\y,-6.5*\y)  node[anchor=west] {\tiny{$p_b$}};
\filldraw[black] (2.3*\y+10*\y,-9.5*\y)  node[anchor=west] {\tiny{$p_a$}};

%Cut On-shell box diagram term 2         

\filldraw[black] (13.5*\y,-4*\y)  node[anchor=west] {\small{$+$}};

\draw[directed] (2.9*\y+13*\y,0.2*\y-5*\y) -- (2.3*\y+13*\y,-0.8*\y-5*\y);
\draw[directed] (2.9*\y+13*\y,0.2*\y-5*\y) -- (2.3*\y+13*\y,2.5*\y-5*\y);
\draw[directed] (2.3*\y+13*\y,-0.8*\y-5*\y) -- (3.8*\y+13*\y,0.7*\y-5*\y);
\draw[directed] (3.8*\y+13*\y,0.7*\y-5*\y) -- (2.3*\y+13*\y,2.5*\y-5*\y);
\draw[directed] (2.9*\y+13*\y,0.2*\y-5*\y) -- (3.8*\y+13*\y,0.7*\y-5*\y);

\filldraw[black] (3.2*\y+13*\y,0*\y-5*\y)  node[anchor=west] {\tiny{$p_a$}};
\filldraw[black] (3.2*\y+13*\y,1.6*\y-5*\y)  node[anchor=west] {\tiny{$p_b$}};
\filldraw[black] (3.2*\y+12*\y,0*\y-5*\y)  node[anchor=west] {\tiny{$B$}};
\filldraw[black] (2.9*\y+12*\y,1.3*\y-5*\y)  node[anchor=west] {\tiny{$B'$}};
\filldraw[black] (3.9*\y+12*\y,0.7*\y-5*\y)  node[anchor=west] {\tiny{$C$}};

\draw[] (17.7*\y,-5.5*\y) -- (17.7*\y,-2.5*\y);
\draw[] (17.7*\y,-5.5*\y) -- (17.4*\y,-5.5*\y);
\draw[] (17.7*\y,-2.5*\y) -- (17.4*\y,-2.5*\y);

%Cut  box diagram term 2     
\filldraw[black] (13.5*\y,-8*\y)  node[anchor=west] {\small{$+$}};

\draw[directed] (1*\y+14*\y,-7.8*\y) -- (1.75*\y+14*\y,-7.5*\y);
\draw[directed] (1.75*\y+14*\y,-8.5*\y) -- (1*\y+14*\y,-8.3*\y);
\draw[directed] (1.75*\y+14*\y,-7.5*\y) -- (1.75*\y+14*\y,-8.5*\y);
\draw[directed] (1.75*\y+14*\y,-8.5*\y) -- (2.45*\y+14*\y,-9.5*\y);
\draw[directed] (1.75*\y+14*\y,-7.5*\y) -- (2.45*\y+14*\y,-6.5*\y);

\filldraw[black] (0.8*\y+14*\y,-7.4*\y)  node[anchor=west] {\tiny{$B'$}};
\filldraw[black] (0.8*\y+14*\y,-8.6*\y)  node[anchor=west] {\tiny{$B$}};
\filldraw[black] (1.7*\y+14*\y,-8*\y)  node[anchor=west] {\tiny{$C$}};
\filldraw[black] (2.3*\y+14*\y,-6.5*\y)  node[anchor=west] {\tiny{$p_b$}};
\filldraw[black] (2.3*\y+14*\y,-9.5*\y)  node[anchor=west] {\tiny{$p_a$}};

\draw[] (17.7*\y,-9.5*\y) -- (17.7*\y,-6.5*\y);
\draw[] (17.7*\y,-9.5*\y) -- (17.4*\y,-9.5*\y);
\draw[] (17.7*\y,-6.5*\y) -- (17.4*\y,-6.5*\y);

\end{tikzpicture}
\end{center}

\caption{Sum over diagrams contributing to the pole at  order $(s-s_0)^{-1}$ and containing on-shell momenta $B$, $B'$ and $C$ (on the left-hand side). The blob $FD^{(j)}$ is any finite tree-level graph having for external states $B$, $B'$ $p_a$ and $p_b$. Since in a tree-level non-allowed $4$-point process the sum over all the finite Feynman diagrams plus the pair of singular ones, in which on-shell bound states propagate, is zero, the sum over the $FD^{(j)}$s yields the two singular diagrams in square brackets on the right-hand side.}
\label{sum_of_order_one_diagrams_written_in_terms_of_order_two_cuts_picture}
\end{figure}

Focusing on the case in figure \ref{sum_of_order_one_diagrams_written_in_terms_of_order_two_cuts_picture}, where the internal on-shell propagators are $\{B, B', C\}$,
we define as integration variables $2 B \cdot \tilde{l} = u$ and $2 B' \cdot \tilde{l} = v$. Then $2 C \cdot \tilde{l}$ is a negative linear combination of them:
$$
2 C \cdot \tilde{l} = -\frac{\Delta_{B'C}}{\Delta_{BB'}} u - \frac{\Delta_{BC}}{\Delta_{BB'}} v.
$$
The sum over all the loop integrals presenting such internal propagating particles can therefore be written as
\begin{equation}
\label{first_expression_for_R1_containing_the_sum_over_the_finite_terms_FDj}
\begin{split}
&\sum_j \frac{1}{(s-s_0)}\frac{1}{8i \Delta_{BB'}}\int \frac{du dv}{(2 \pi)^2} \frac{1}{u+i\epsilon} \  \frac{1}{v+i\epsilon} \ \frac{1}{\frac{dC^2}{ds}\Bigr|_{B,B'}-\frac{\Delta_{B'C}}{\Delta_{BB'}} u - \frac{\Delta_{BC}}{\Delta_{BB'}} v  +i\epsilon} \ FD^{(j)}_{abB B'}(s_0) \times 2\\
&=\frac{1}{(s-s_0)}\frac{i}{8 \Delta_{BB'}} \frac{1}{\frac{dC^2}{ds}\Bigr|_{B,B'}} \sum_j FD^{(j)}_{abB B'}(s_0) \times 2.
\end{split}
\end{equation}
On the right-hand side of the equality, we have simply computed the integral by closing the $u$ and $v$ contours in the lower half complex plane. The multiplicity factor $2$ at the end of the equation is because, for any $j$, there are exactly two copies of such diagrams, connected by inverting the direction of the arrows and rotating the on-shell parallelogram by $180^\circ$. 
The term $FD^{(j)}_{abB B'}(s_0)$ is any tree-level diagram having as external states $p_a$, $p_b$, $B$ and $B'$ which is finite at the pole position $s=s_0$ in such a way to  not generate higher-order singularities. At this point, we need to sum over all of these finite tree-level diagrams contributing to the scattering of  $p_a$, $p_b$, $B$, $B'$.  In doing this an important property of the tree level integrability of the theory comes to our aid. Since the process involving the scattering of $p_a$, $p_b$, $B$, $B'$ 
is inelastic, and therefore forbidden, the sum over all the Feynman diagrams contributing to such process needs to be null. Such a sum contains all the diagrams that are finite at the value $s=s_0$, i.e.\  the terms $FD^{(j)}_{abB B'}$ in \eqref{first_expression_for_R1_containing_the_sum_over_the_finite_terms_FDj}, plus two singular contributions in which the on-shell bound states $C$ and $C'$ propagate.
Since the total sum is null, the term $\sum_j FD^{(j)}_{abB B'}(s_0)$ is equal to minus the two diagrams in which $C$ and $C'$ propagate
and the expression in \eqref{first_expression_for_R1_containing_the_sum_over_the_finite_terms_FDj}, after having inserted the $3$-point couplings, can be written as
\begin{equation}
\label{first_expression_for_R1}
R^{(1)}= - \frac{1}{(s-s_0)}\frac{i}{8 \Delta_{BB'}} \frac{C_{aB \bar{C}} C_{Cb \bar{B'}}}{\frac{dC^2}{ds}\Bigr|_{B,B'}} \Bigl[ \frac{C_{\bar{B} \bar{b} C'} C_{\bar{C'} B' \bar{a}}}{C'^2-m^2_{C'}} + \frac{C_{B' \bar{b} \bar{C}} C_{C \bar{B} \bar{a}}}{C^2-m^2_C}  \Bigr] \times 2.
\end{equation}
A pictorial representation of the identity is shown in figure \ref{sum_of_order_one_diagrams_written_in_terms_of_order_two_cuts_picture}. Even though the expression in square brackets in \eqref{first_expression_for_R1} may seem to contain a further pole, since when $s=s_0$ the $C$ and $C'$ momenta are on-shell, as we explained in section \ref{section_tree_level_and_bootstrap} that is not the case. The pair of diagrams indeed have singularities that cancel each other, and what remains is a finite contribution given by Taylor-expanding $C^2$ and $C'^2$ to second order in $(s-s_0)$ 
\begin{equation}
\label{second_expression_for_R1}
R^{(1)}= \frac{1}{(s-s_0)}\frac{i}{16 \Delta_{BB'}} \frac{C_{aB \bar{C}} C_{Cb \bar{B'}}}{\frac{dC^2}{ds}} \Bigl[ C_{\bar{B} \bar{b} C'} C_{\bar{C'} B' \bar{a}} \frac{\frac{d^2 C'^2}{ds^2}}{\bigl( \frac{d C'^2}{ds} \bigr)^2} + C_{B' \bar{b} C} C_{\bar{C} \bar{B} \bar{a}}  \frac{\frac{d^2 C^2}{ds^2}}{\bigl( \frac{d C^2}{ds} \bigr)^2}   \Bigr] \times 2.
\end{equation}
In the expression above the derivative needs to be performed keeping the lengths of the momenta $p_a$, $p_b$, $B$ and $B'$ fixed on their respective on-shell values. The products of the $3$-point couplings are the same as those appearing in the  second-order pole network: they are always positive and after some computations, it is possible to check that
\begin{equation}
\label{third_expression_for_R1}
R^{(1)}= i \Bigl( \frac{\beta}{\sqrt{h}} \Bigr)^4 \frac{2 \Delta_{ab}}{s-s_0} \biggl[ p_a^2 \frac{\Delta_{B'C} \Delta_{BC'}}{\Delta_{BC}\Delta_{B'C'}}+p_b^2 \frac{\Delta_{BC} \Delta_{B'C'}}{\Delta_{B'C}\Delta_{BC'}} + 2 p_a \cdot p_b \biggr] \times 2.
\end{equation}
A similar computation for the set of diagrams with internal on-shell propagators $\{C, C', B\}$ yields
\begin{equation}
\label{second_expression_for_R2}
R^{(2)}=i \Bigl( \frac{\beta}{\sqrt{h}} \Bigr)^4 \frac{2 \Delta_{ab}}{s-s_0} \biggl[ -p_a^2 \frac{\Delta_{B'C} \Delta_{BC'}}{\Delta_{BC}\Delta_{B'C'}}-p_b^2 \frac{\Delta_{BC} \Delta_{B'C'}}{\Delta_{B'C}\Delta_{BC'}} + 2 p_a \cdot p_b \biggr] \times 2.
\end{equation}
Summing~\eqref{third_expression_for_R1} and~\eqref{second_expression_for_R2} we obtain
\begin{equation}
\label{second_expression_for_R1_plus_R2}
R^{(1)}+R^{(2)}=16i \Bigl( \frac{\beta}{\sqrt{h}} \Bigr)^4 \frac{ \Delta_{ab}}{s-s_0}  p_a \cdot p_b.
\end{equation}

If there are $N$ disjoint networks of the form in figure~\ref{allowed_one_loop_box_network_on_shell_and_F_Diagram_description} the computation needs to be repeated $N$ times so that the sum of all the diagrams contributing to the order one pole in~\eqref{Expansion_of_S_around_an_arbitrary_2N_order_pole} is
\begin{equation}
\label{second_expression_for_R1_plus_R2_with_N}
N \Bigl[16i \Bigl( \frac{\beta}{\sqrt{h}} \Bigr)^4 \frac{ \Delta_{ab}}{s-s_0}  p_a \cdot p_b \Bigr].
\end{equation}
Summing~\eqref{expansion_of_N_copies_of_the_network} and~\eqref{second_expression_for_R1_plus_R2_with_N},  the singular part of the amplitude at the order $\beta^4$ is
\begin{equation}
\label{full_second_order_pole_amplitude_in_perturbation_theory}
M^{(\text{sing},\beta^4)}_{ab}= N\Bigl[ -32 i \Bigl( \frac{\beta}{\sqrt{h}} \Bigr)^4 \frac{\Delta_{ab}^3}{(s-s_0)^2}+
8i \Bigl( \frac{\beta}{\sqrt{h}} \Bigr)^4 \frac{\Delta_{ab}}{s-s_0} p_a \cdot p_b\Bigr].
\end{equation}
Before comparing with the bootstrapped S-matrix we should
divide this expression by the proper normalisation factor coming from the Dirac delta function of overall energy-momentum conservation, to obtain
$$
S^{(\text{sing},\beta^4)}_{ab}(\theta)=\frac{1}{4m_a m_b \sinh \theta} M^{(\text{sing},\beta^4)}_{ab} = \frac{1}{8i \Delta_{ab}}  \Bigl[1 +i \operatorname{cotan} \theta_0 (\theta-i\theta_0) +\ldots \Bigr] M^{(\text{sing},\beta^4)}_{ab}.
$$
Noting that $p_a \cdot p_b= m_a m_b \cosh i \theta_0=2 \Delta_{ab} \operatorname{cotan} \theta_0$, 
and writing the Mandelstam variable $s$ in terms of the rapidity difference $\theta$ so that 
$$
\frac{1}{s-s_0}= \frac{1}{2 m_a m_b \bigl(\cosh \theta - \cosh i \theta_0 \bigr)} = \frac{1}{4 i \Delta_{ab} \bigl(\theta - i \theta_0 \bigr)} \Bigl[ 1  +\frac{i}{2} \operatorname{cotan} \theta_0 \bigl(\theta - i \theta_0\bigr)
+\dots\Bigr],
$$
the simple pole terms in $(\theta-i\theta_0)$ cancel and we end up with
\begin{equation}
\label{full_second_order_pole_S_matrix_in_perturbation_theory}
S^{(\text{sing},\beta^4)}_{ab}(\theta)=   \Bigl( \frac{\beta}{\sqrt{2h}} \Bigr)^4 \frac{N}{(\theta-i \theta_0)^2}\,.
\end{equation}
This implies the values  $a_2=N$ and $b_1=0$ for the order $\beta^4$ coefficients
of the expansion~\eqref{Expansion_of_S_around_an_arbitrary_2N_order_pole}. They have a universal form, not depending on the simply-laced theory under consideration, and match perfectly with the bootstrapped result \eqref{S_matrix_expansion_obtained_using_bootstrap_on_the_pole}.
\begin{figure}

\begin{center}
\begin{tikzpicture}
\tikzmath{\y=0.5;}

%Flip lines network 1
\draw[] (1*\y,1*\y) -- (2*\y,0*\y);
\draw[] (-5.5*\y,-3.5*\y) -- (-4.5*\y,-4.5*\y);
\draw[] (-4.5*\y,1*\y) -- (-5.5*\y,0*\y);
\draw[] (2*\y,-3.5*\y) -- (1*\y,-4.5*\y);

%Network 1, diagram 1
\draw[] (-0.522642*\y-1.868096*\y,2.45884*\y-1.682042*\y) -- (-3.09017*\y,0*\y);

\draw[] (-0.522642*\y-1.868096*\y,2.45884*\y-1.682042*\y) -- (-3.09017*\y+-0.522642*\y,0*\y+2.45884*\y);
\draw[] (0*\y+0*\y, 0*\y+0*\y) -- (-0.522642*\y+0*\y,2.45884*\y+0*\y);
\draw[] (-3.09017*\y+-0.522642*\y,0*\y+2.45884*\y) -- (-3.09017*\y+0*\y,0*\y+0*\y);
\draw[] (-0.522642*\y-1.868096*\y,2.45884*\y-1.682042*\y) -- (-0.522642*\y+0*\y,2.45884*\y+0*\y);
\draw[] (-0.522642*\y-1.868096*\y,2.45884*\y-1.682042*\y) -- (0*\y+0*\y, 0*\y+0*\y);
\draw[] (0*\y+0*\y, 0*\y+0*\y) -- (-3.09017*\y+0*\y,0*\y+0*\y);
\draw[] (0*\y+-0.522642*\y, 0*\y+2.45884*\y) -- (-3.09017*\y+-0.522642*\y,0*\y+2.45884*\y);

\filldraw[] (-2.6*\y,1.9*\y)  node[anchor=west] {\tiny{$\bullet$}};
\filldraw[] (-1.3*\y,1*\y)  node[anchor=west] {\tiny{$\bullet$}};

\filldraw[] (-0.4*\y,1.3*\y)  node[anchor=west] {\tiny{$4$}};
\filldraw[] (-4.05*\y,1.3*\y)  node[anchor=west] {\tiny{$4$}};
\filldraw[] (-1.8*\y,1.7*\y)  node[anchor=west] {\tiny{$4$}};
\filldraw[] (-1.8*\y,0.7*\y)  node[anchor=west] {\tiny{$4$}};
\filldraw[] (-2.3*\y,2.7*\y)  node[anchor=west] {\tiny{$5$}};
\filldraw[] (-2.3*\y,-0.22*\y)  node[anchor=west] {\tiny{$5$}};
\filldraw[] (-3.2*\y,1.7*\y)  node[anchor=west] {\tiny{$3$}};
\filldraw[] (-3.2*\y,0.6*\y)  node[anchor=west] {\tiny{$1$}};

%Network 1, diagram 2
\draw[] (-0.522642*\y-1.868096*\y+4.5*\y,2.45884*\y-1.682042*\y-3*\y) -- (-3.09017*\y+4.5*\y,0*\y-3*\y);
\draw[] (-0.522642*\y+4.5*\y,2.45884*\y+0*\y-3*\y) -- (-0.699432*\y+-0.522642*\y+4.5*\y,-0.776798*\y+2.45884*\y-3*\y);
\draw[] (-0.522642*\y-1.868096*\y+4.5*\y,2.45884*\y-1.682042*\y-3*\y) -- (-3.09017*\y+-0.522642*\y+4.5*\y,0*\y+2.45884*\y-3*\y);
\draw[] (-0.699432*\y+-0.522642*\y+4.5*\y,-0.776798*\y+2.45884*\y-3*\y) -- (0*\y+4.5*\y, -3*\y);
\draw[] (+4.5*\y, -3*\y) -- (-0.522642*\y+4.5*\y,2.45884*\y-3*\y);
\draw[] (-3.09017*\y+-0.522642*\y+4.5*\y,0*\y+2.45884*\y-3*\y) -- (-3.09017*\y+4.5*\y,-3*\y);
\draw[] (-0.522642*\y-1.868096*\y+4.5*\y,2.45884*\y-1.682042*\y-3*\y) -- (+4.5*\y, -3*\y);
\draw[] (-0.699432*\y+-0.522642*\y+4.5*\y,-0.776798*\y+2.45884*\y-3*\y) -- (-3.09017*\y+-0.522642*\y+4.5*\y,0*\y+2.45884*\y-3*\y);
\draw[] (+4.5*\y, -3*\y) -- (-3.09017*\y+4.5*\y,-3*\y);
\draw[] (0*\y+-0.522642*\y+4.5*\y, 0*\y+2.45884*\y-3*\y) -- (-3.09017*\y+-0.522642*\y+4.5*\y,0*\y+2.45884*\y-3*\y);

\filldraw[] (2.2*\y,-1.8*\y)  node[anchor=west] {\tiny{$\bullet$}};

\filldraw[] (-0.4*\y+4.5*\y,1.3*\y-3*\y)  node[anchor=west] {\tiny{$4$}};
\filldraw[] (-4.0*\y+4.5*\y,1.3*\y-3*\y)  node[anchor=west] {\tiny{$4$}};
\filldraw[] (-2.7*\y+4.5*\y,1.8*\y-3*\y)  node[anchor=west] {\tiny{$4$}};
\filldraw[] (-1.8*\y+4.5*\y,0.7*\y-3*\y)  node[anchor=west] {\tiny{$4$}};
\filldraw[] (-2.3*\y+4.5*\y,2.7*\y-3*\y)  node[anchor=west] {\tiny{$5$}};
\filldraw[] (-2.3*\y+4.5*\y,-0.2*\y-3*\y)  node[anchor=west] {\tiny{$5$}};
\filldraw[] (-3.3*\y+4.5*\y,1.4*\y-3*\y)  node[anchor=west] {\tiny{$3$}};
\filldraw[] (-1.4*\y+4.5*\y,1.3*\y-3*\y)  node[anchor=west] {\tiny{$3$}};
\filldraw[] (-3.2*\y+4.5*\y,0.6*\y-3*\y)  node[anchor=west] {\tiny{$1$}};
\filldraw[] (-1.2*\y+4.5*\y,2.2*\y-3*\y)  node[anchor=west] {\tiny{$1$}};

%Network 1, diagram 3

\draw[] (-0.522642*\y+0*\y,2.45884*\y+0*\y-6*\y) -- (-0.699432*\y+-0.522642*\y,-0.776798*\y+2.45884*\y-6*\y);
\draw[] (-0.699432*\y+-0.522642*\y,-0.776798*\y+2.45884*\y-6*\y) -- (0*\y, -6*\y);
\draw[] (-0.699432*\y+-0.522642*\y,-0.776798*\y+2.45884*\y-6*\y) -- (-3.09017*\y+-0.522642*\y,0*\y+2.45884*\y-6*\y);
\draw[] (0*\y, -6*\y) -- (-0.522642*\y,2.45884*\y+0*\y-6*\y);
\draw[] (-3.09017*\y+-0.522642*\y,0*\y+2.45884*\y-6*\y) -- (-3.09017*\y,-6*\y);
\draw[] (-3.09017*\y,-6*\y) -- (-0.699432*\y+-0.522642*\y,-0.776798*\y+2.45884*\y-6*\y);
\draw[] (0*\y, -6*\y) -- (-3.09017*\y,-6*\y);
\draw[] (0*\y+-0.522642*\y, 0*\y+2.45884*\y-6*\y) -- (-3.09017*\y+-0.522642*\y,0*\y+2.45884*\y-6*\y);

\filldraw[] (-1.8*\y,-5.3*\y)  node[anchor=west] {\tiny{$\bullet$}};
\filldraw[] (-3.1*\y,-4.5*\y)  node[anchor=west] {\tiny{$\bullet$}};

\filldraw[] (-0.4*\y,1.3*\y-6*\y)  node[anchor=west] {\tiny{$4$}};
\filldraw[] (-4.05*\y,1.3*\y-6*\y)  node[anchor=west] {\tiny{$4$}};
\filldraw[] (-2.5*\y,1.7*\y-6*\y)  node[anchor=west] {\tiny{$4$}};
\filldraw[] (-2.3*\y,2.7*\y-6*\y)  node[anchor=west] {\tiny{$5$}};
\filldraw[] (-2.3*\y,-0.2*\y-6*\y)  node[anchor=west] {\tiny{$5$}};
\filldraw[] (-1.3*\y,1.2*\y-6*\y)  node[anchor=west] {\tiny{$3$}};
\filldraw[] (-3.0*\y,0.8*\y-6*\y)  node[anchor=west] {\tiny{$4$}};
\filldraw[] (-1.2*\y,2.2*\y-6*\y)  node[anchor=west] {\tiny{$1$}};

%Network 1, diagram 4

\draw[] (-0.522642*\y+0*\y-4.5*\y,2.45884*\y+0*\y-3*\y) -- (-0.699432*\y+-0.522642*\y-4.5*\y,-0.776798*\y+2.45884*\y-3*\y);
\draw[] (-0.522642*\y-1.868096*\y-4.5*\y,2.45884*\y-1.682042*\y-3*\y) -- (-3.09017*\y-4.5*\y,0*\y-3*\y);
\draw[] (-0.699432*\y+-0.522642*\y-4.5*\y,-0.776798*\y+2.45884*\y-3*\y) -- (-4.5*\y, -3*\y);
\draw[] (-0.522642*\y-1.868096*\y-4.5*\y,2.45884*\y-1.682042*\y-3*\y) -- (-3.09017*\y+-0.522642*\y-4.5*\y,0*\y+2.45884*\y-3*\y);
\draw[] (-4.5*\y, -3*\y) -- (-0.522642*\y-4.5*\y,2.45884*\y+0*\y-3*\y);
\draw[] (-3.09017*\y+-0.522642*\y-4.5*\y,0*\y+2.45884*\y-3*\y) -- (-3.09017*\y-4.5*\y,-3*\y);
\draw[] (-3.09017*\y-4.5*\y,-3*\y) -- (-0.699432*\y+-0.522642*\y-4.5*\y,-0.776798*\y+2.45884*\y-3*\y);
\draw[] (-0.522642*\y-1.868096*\y-4.5*\y,2.45884*\y-1.682042*\y-3*\y) -- (-0.522642*\y-4.5*\y,2.45884*\y-3*\y);
\draw[] (-4.5*\y, -3*\y) -- (-3.09017*\y-4.5*\y,-3*\y);
\draw[] (0*\y+-0.522642*\y-4.5*\y, 0*\y+2.45884*\y-3*\y) -- (-3.09017*\y+-0.522642*\y-4.5*\y,0*\y+2.45884*\y-3*\y);

\filldraw[] (-6.4*\y,-2.3*\y)  node[anchor=west] {\tiny{$\bullet$}};
\filldraw[] (-6.7*\y,-1.0*\y)  node[anchor=west] {\tiny{$\bullet$}};

\filldraw[] (-0.4*\y-4.5*\y,1.3*\y-3*\y)  node[anchor=west] {\tiny{$4$}};
\filldraw[] (-4.05*\y-4.5*\y,1.3*\y-3*\y)  node[anchor=west] {\tiny{$4$}};
\filldraw[] (-2.5*\y-4.5*\y,1.4*\y-3*\y)  node[anchor=west] {\tiny{$4$}};
\filldraw[] (-2.3*\y-4.5*\y,2.7*\y-3*\y)  node[anchor=west] {\tiny{$5$}};
\filldraw[] (-2.3*\y-4.5*\y,-0.2*\y-3*\y)  node[anchor=west] {\tiny{$5$}};
\filldraw[] (-1.3*\y-4.5*\y,1.2*\y-3*\y)  node[anchor=west] {\tiny{$3$}};
\filldraw[] (-2.4*\y-4.5*\y,0.7*\y-3*\y)  node[anchor=west] {\tiny{$4$}};
\filldraw[] (-1.1*\y-4.5*\y,1.9*\y-3*\y)  node[anchor=west] {\tiny{$1$}};
\filldraw[] (-3.2*\y-4.5*\y,0.6*\y-3*\y)  node[anchor=west] {\tiny{$1$}};
\filldraw[] (-3.2*\y-4.5*\y,1.7*\y-3*\y)  node[anchor=west] {\tiny{$3$}};

%Flip lines network 2
\draw[] (16*\y,1*\y) -- (17*\y,0*\y);
\draw[] (9.5*\y,-3.5*\y) -- (10.5*\y,-4.5*\y);
\draw[] (10.5*\y,1*\y) -- (9.5*\y,0*\y);
\draw[] (17*\y,-3.5*\y) -- (16*\y,-4.5*\y);

%Network 2, diagram 1

\draw[] (-3.09017*\y+-0.522642*\y+15*\y, 0*\y+2.45884*\y) -- (-2.39074*\y+-0.522642*\y+15*\y,-0.776798*\y+2.45884*\y);
\draw[] (-2.39074*\y+-0.522642*\y+15*\y,-0.776798*\y+2.45884*\y) -- (-3.09017*\y+15*\y,0*\y+0*\y);
\draw[] (+15*\y, 0*\y+0*\y) -- (-0.522642*\y+15*\y,2.45884*\y+0*\y);
\draw[] (0*\y+-0.522642*\y+15*\y, 0*\y+2.45884*\y) -- (-2.39074*\y+-0.522642*\y+15*\y,-0.776798*\y+2.45884*\y);
\draw[] (-3.09017*\y+-0.522642*\y+15*\y,0*\y+2.45884*\y) -- (-3.09017*\y+15*\y,0*\y+0*\y);
\draw[] (+15*\y, 0*\y+0*\y) -- (-3.09017*\y+15*\y,0*\y+0*\y);
\draw[] (0*\y+-0.522642*\y+15*\y, 0*\y+2.45884*\y) -- (-3.09017*\y+-0.522642*\y+15*\y,0*\y+2.45884*\y);
\draw[] (-2.39074*\y+-0.522642*\y+15*\y,-0.776798*\y+2.45884*\y) -- (+15*\y, 0*\y+0*\y);

\filldraw[] (13.5*\y,1.3*\y)  node[anchor=west] {\tiny{$\bullet$}};
\filldraw[] (12.5*\y,0.4*\y)  node[anchor=west] {\tiny{$\bullet$}};

\filldraw[] (-0.4*\y+15*\y,1.3*\y)  node[anchor=west] {\tiny{$4$}};
\filldraw[] (-4.05*\y+15*\y,1.3*\y)  node[anchor=west] {\tiny{$4$}};
\filldraw[] (-1.8*\y+15*\y,1.9*\y)  node[anchor=west] {\tiny{$4$}};
\filldraw[] (-1.8*\y+15*\y,0.8*\y)  node[anchor=west] {\tiny{$6$}};
\filldraw[] (-2.3*\y+15*\y,2.7*\y)  node[anchor=west] {\tiny{$5$}};
\filldraw[] (-2.3*\y+15*\y,-0.2*\y)  node[anchor=west] {\tiny{$5$}};
\filldraw[] (-3.2*\y+15*\y,1.0*\y)  node[anchor=west] {\tiny{$2$}};
\filldraw[] (-3.5*\y+15*\y,2.2*\y)  node[anchor=west] {\tiny{$1$}};

%Network 2, diagram 2

\draw[] (-3.09017*\y+-0.522642*\y+19.5*\y, 0*\y+2.45884*\y-3*\y) -- (-2.39074*\y+-0.522642*\y+19.5*\y,-0.776798*\y+2.45884*\y-3*\y);
\draw[] (-3.09017*\y+2.39074*\y+19.5*\y, 0.776798*\y-3*\y) -- (+19.5*\y,0*\y-3*\y);
\draw[] (-2.39074*\y+-0.522642*\y+19.5*\y,-0.776798*\y+2.45884*\y-3*\y) -- (-3.09017*\y+19.5*\y,0*\y+0*\y-3*\y);
\draw[] (-3.09017*\y+2.39074*\y+19.5*\y, 0.776798*\y-3*\y) -- (0*\y+-0.522642*\y+19.5*\y, 0*\y+2.45884*\y-3*\y);
\draw[] (+19.5*\y, -3*\y) -- (-0.522642*\y+19.5*\y,2.45884*\y-3*\y);
\draw[] (-3.09017*\y+-0.522642*\y+19.5*\y,+2.45884*\y-3*\y) -- (-3.09017*\y+19.5*\y,-3*\y);
\draw[] (+19.5*\y, -3*\y) -- (-3.09017*\y+19.5*\y,-3*\y);
\draw[] (0*\y+-0.522642*\y+19.5*\y, +2.45884*\y-3*\y) -- (-3.09017*\y+-0.522642*\y+19.5*\y,+2.45884*\y-3*\y);
\draw[] (-2.39074*\y+-0.522642*\y+19.5*\y,-0.776798*\y+2.45884*\y-3*\y) -- (+19.5*\y, -3*\y);
\draw[] (-3.09017*\y+2.39074*\y+19.5*\y, 0.776798*\y-3*\y) -- (-3.09017*\y+-0.522642*\y+19.5*\y,+2.45884*\y-3*\y);

\filldraw[] (17.5*\y,-1*\y)  node[anchor=west] {\tiny{$\bullet$}};
\filldraw[] (17.1*\y,-2.5*\y)  node[anchor=west] {\tiny{$\bullet$}};

\filldraw[] (-0.4*\y+19.5*\y,1.3*\y-3*\y)  node[anchor=west] {\tiny{$4$}};
\filldraw[] (-4.05*\y+19.5*\y,1.3*\y-3*\y)  node[anchor=west] {\tiny{$4$}};
\filldraw[] (-1.2*\y+19.5*\y,1.9*\y-3*\y)  node[anchor=west] {\tiny{$2$}};
\filldraw[] (-1.8*\y+19.5*\y,0.8*\y-3*\y)  node[anchor=west] {\tiny{$6$}};
\filldraw[] (-2.3*\y+19.5*\y,2.7*\y-3*\y)  node[anchor=west] {\tiny{$5$}};
\filldraw[] (-2.3*\y+19.5*\y,-0.2*\y-3*\y)  node[anchor=west] {\tiny{$5$}};
\filldraw[] (-3.2*\y+19.5*\y,1.0*\y-3*\y)  node[anchor=west] {\tiny{$2$}};
\filldraw[] (-3.65*\y+19.5*\y,1.8*\y-3*\y)  node[anchor=west] {\tiny{$1$}};
\filldraw[] (-2.6*\y+19.8*\y,1.5*\y-3*\y)  node[anchor=west] {\tiny{$6$}};
\filldraw[] (-0.75*\y+19.5*\y,0.7*\y-3*\y)  node[anchor=west] {\tiny{$1$}};

%Network 2, diagram 3

\draw[] (-3.09017*\y+2.39074*\y+15*\y, 0.776798*\y-6*\y) -- (+15*\y,0*\y-6*\y);
\draw[] (-3.09017*\y+2.39074*\y+15*\y, 0.776798*\y-6*\y) -- (0*\y+-0.522642*\y+15*\y, 0*\y+2.45884*\y-6*\y);
\draw[] (+15*\y, -6*\y) -- (-0.522642*\y+15*\y,2.45884*\y-6*\y);
\draw[] (-3.09017*\y+-0.522642*\y+15*\y,+2.45884*\y-6*\y) -- (-3.09017*\y+15*\y,-6*\y);
\draw[] (-3.09017*\y+2.39074*\y+15*\y, 0.776798*\y-6*\y) -- (-3.09017*\y+15*\y,-6*\y);
\draw[] (+15*\y, -6*\y) -- (-3.09017*\y+15*\y,-6*\y);
\draw[] (0*\y+-0.522642*\y+15*\y, +2.45884*\y-6*\y) -- (-3.09017*\y+-0.522642*\y+15*\y,+2.45884*\y-6*\y);
\draw[] (-3.09017*\y+2.39074*\y+15*\y, 0.776798*\y-6*\y) -- (-3.09017*\y+-0.522642*\y+15*\y,+2.45884*\y-6*\y);

\filldraw[] (12.2*\y,-5*\y)  node[anchor=west] {\tiny{$\bullet$}};
\filldraw[] (13*\y,-4*\y)  node[anchor=west] {\tiny{$\bullet$}};

\filldraw[] (-0.4*\y+15*\y,1.3*\y-6*\y)  node[anchor=west] {\tiny{$4$}};
\filldraw[] (-4.05*\y+15*\y,1.3*\y-6*\y)  node[anchor=west] {\tiny{$4$}};
\filldraw[] (-1.2*\y+15*\y,1.9*\y-6*\y)  node[anchor=west] {\tiny{$2$}};
\filldraw[] (-2*\y+15*\y,0.7*\y-6*\y)  node[anchor=west] {\tiny{$4$}};
\filldraw[] (-2.3*\y+15*\y,2.7*\y-6*\y)  node[anchor=west] {\tiny{$5$}};
\filldraw[] (-2.3*\y+15*\y,-0.2*\y-6*\y)  node[anchor=west] {\tiny{$5$}};
\filldraw[] (-3.0*\y+15*\y,1.6*\y-6*\y)  node[anchor=west] {\tiny{$6$}};
\filldraw[] (-0.75*\y+15*\y,0.7*\y-6*\y)  node[anchor=west] {\tiny{$1$}};

%Network 2, diagram 4

\draw[] (-3.09017*\y+-0.522642*\y+10.5*\y, 0*\y+2.45884*\y-3*\y) -- (-2.39074*\y+-0.522642*\y+10.5*\y,-0.776798*\y+2.45884*\y-3*\y);
\draw[] (-3.09017*\y+2.39074*\y+10.5*\y, 0.776798*\y-3*\y) -- (+10.5*\y,0*\y-3*\y);
\draw[] (-3.09017*\y+2.39074*\y+10.5*\y, 0.776798*\y-3*\y) -- (0*\y+-0.522642*\y+10.5*\y, 0*\y+2.45884*\y-3*\y);
\draw[] (-2.39074*\y+-0.522642*\y+10.5*\y,-0.776798*\y+2.45884*\y-3*\y) -- (-3.09017*\y+10.5*\y,0*\y+0*\y-3*\y);
\draw[] (+10.5*\y, -3*\y) -- (-0.522642*\y+10.5*\y,2.45884*\y-3*\y);
\draw[] (-3.09017*\y+-0.522642*\y+10.5*\y,+2.45884*\y-3*\y) -- (-3.09017*\y+10.5*\y,-3*\y);
\draw[] (-3.09017*\y+2.39074*\y+10.5*\y, 0.776798*\y-3*\y) -- (-3.09017*\y+10.5*\y,-3*\y);
\draw[] (0*\y+-0.522642*\y+10.5*\y, 0*\y+2.45884*\y-3*\y) -- (-2.39074*\y+-0.522642*\y+10.5*\y,-0.776798*\y+2.45884*\y-3*\y);
\draw[] (+10.5*\y, -3*\y) -- (-3.09017*\y+10.5*\y,-3*\y);
\draw[] (0*\y+-0.522642*\y+10.5*\y, +2.45884*\y-3*\y) -- (-3.09017*\y+-0.522642*\y+10.5*\y,+2.45884*\y-3*\y);

\filldraw[] (8.5*\y,-1.8*\y)  node[anchor=west] {\tiny{$\bullet$}};

\filldraw[] (-0.4*\y+10.5*\y,1.3*\y-3*\y)  node[anchor=west] {\tiny{$4$}};
\filldraw[] (-4.05*\y+10.5*\y,1.3*\y-3*\y)  node[anchor=west] {\tiny{$4$}};
\filldraw[] (-1.2*\y+10.5*\y,1.7*\y-3*\y)  node[anchor=west] {\tiny{$2$}};
\filldraw[] (-2*\y+10.5*\y,0.6*\y-3*\y)  node[anchor=west] {\tiny{$4$}};
\filldraw[] (-2.3*\y+10.5*\y,2.7*\y-3*\y)  node[anchor=west] {\tiny{$5$}};
\filldraw[] (-2.3*\y+10.5*\y,-0.2*\y-3*\y)  node[anchor=west] {\tiny{$5$}};
\filldraw[] (-2.3*\y+10.5*\y,1.7*\y-3*\y)  node[anchor=west] {\tiny{$4$}};
\filldraw[] (-0.75*\y+10.5*\y,0.7*\y-3*\y)  node[anchor=west] {\tiny{$1$}};
\filldraw[] (-3.2*\y+10.5*\y,1.2*\y-3*\y)  node[anchor=west] {\tiny{$2$}};
\filldraw[] (-3.5*\y+10.5*\y,2.2*\y-3*\y)  node[anchor=west] {\tiny{$1$}};

\end{tikzpicture}
\end{center}
\caption{Two networks which contribute to a $4$th-order pole in the $e_8^{(1)}$ affine Toda theory. The pole is at $\theta_0= \frac{13 \pi}{30}$ where the S-matrix element is $S_{45}(\theta) \sim \{12\}^2 \{14\}^2$ in the notation of appendix~\ref{Appendix_S_matrix_expansion}. Only the lightest
six of the eight particles present in the model contribute to these networks; they are labelled in increasing order of mass $m_1<m_2<\ldots<m_6$. Bullets have been added inside the triangles and parallelograms that can be subtiled.}  
\label{Double_copy_of_the_network_in_4th_order_pole_e8}
\end{figure}
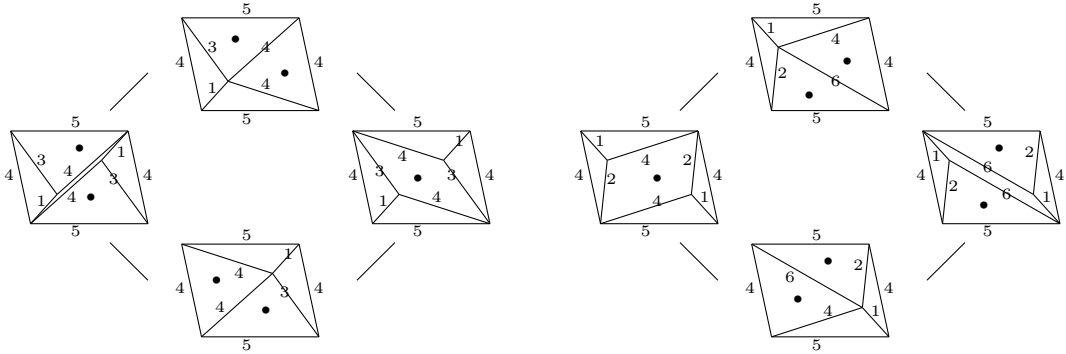

To illustrate some of the issues that arise when extending this result to higher orders, figure \ref{Double_copy_of_the_network_in_4th_order_pole_e8}  shows the two copies of the network discussed above that appear in the $e_8^{(1)}$ affine Toda model at one particular $4^{\text{th}}$-order pole. Since in this case $N=2$ there are two networks, both contributing to the double and single pole coefficients at order $\beta^4$.
However the two networks, that appear separated at one loop, are connected if we look at higher loop orders. 
Some of the tiles composing the parallelogram corresponding to the on-shell process can be additionally tiled into sub-pieces, and the networks that look disjoint at one loop can be connected by flipping internal propagators. 
This means that even finding all the on-shell diagrams contributing to the coefficient $a_4$ in the Laurent expansion of the S-matrix is a very difficult task. To achieve the goal, we should further tile all the constituents of figure~\ref{Double_copy_of_the_network_in_4th_order_pole_e8} containing a bullet. The Feynman diagrams generated in this way  contain three loops, and from them, by flipping propagators, we should try to generate the full network of singular diagrams. The total number of these singular diagrams is very large and is connected with the concept of the depth of a diagram, introduced in~\cite{Braden:1990wx}. 
Investigating the higher loop structure of the diagrams is necessary to capture all the coefficients of the Laurent expansion, up to  order $(\theta-i\theta_0)^{-2N}$ in formula~\eqref{Expansion_of_S_around_an_arbitrary_2N_order_pole}, but
we leave further discussion of this point for a companion publication~\cite{Second_loop_paper_sagex}.

It is also important to remark that is possible to add further vertices to the on-shell diagrams in figure~\ref{Double_copy_of_the_network_in_4th_order_pole_e8} which do not increase the order of the pole. This corresponds to evaluating higher powers of $\beta$ contributing to the coefficient at order $p=2$ in~\eqref{Expansion_of_S_around_an_arbitrary_2N_order_pole}.

\section{Conclusions}
\label{The_Conclusions}
The mechanism responsible for the generation of first- and second-order singularities present in the Laurent expansion of the S-matrix around a generic even-order pole has been investigated up to  order $\beta^4$, which is one-loop in perturbation theory, for the entire class of ADE affine Toda models.

Once all relevant diagrams were summed, the results we found became universal, not depending on the particular theory or process considered.
In all  cases we have shown that the loop integrals in the neighbourhood of the Landau singularity can be broken into the product of tree-level graphs and the expected pole residues are recovered by exploiting the tree-level integrability properties of the model. The additional multiplicity factor $N$ found in the bootstrapped formula \eqref{S_matrix_expansion_obtained_using_bootstrap_on_the_pole} at a $2N^{\rm th}$ order singularity is reproduced in perturbation theory noting that at the value of the rapidity $\theta=i\theta_0$ there are exactly $N$ copies of the network in figure \ref{allowed_one_loop_box_network_on_shell_and_F_Diagram_description} contributing to the final result. Through our study we have shown that the low-order coefficients of the Laurent expansion of the S-matrix at a general even-order pole (as given in equation~\eqref{Expansion_of_S_around_an_arbitrary_2N_order_pole}) are $a_1=b_1=0$ and $a_2=N$. 

The knowledge of all the coefficients of the Laurent expansion at the poles is a necessary ingredient to reconstruct the scattering using the so-called dispersion relation, that connects the S-matrix evaluated at a generic point of the Mandelstam variable $s$ to its values at the poles and cuts. A further study of higher-order singularities and how to reduce bigger networks of Feynman diagrams contributing to Landau singularities to simple expressions will be the subject of a companion paper~\cite{Second_loop_paper_sagex}.

The axiomatic procedure used in the past to determine the S-matrices of a variety of (1+1)-dimensional integrable theories, though passing many non-trivial tests, relies on a sequence of conjectures, and its connection with the usual perturbative approach remains to be fully understood. For example, the mechanism by which Feynman diagrams contributing to non-elastic processes  sum to zero is still unknown, apart from at tree level \cite{Patrick_Davide_paper}, as is a full understanding of the emergence of higher-order singularities. It would be interesting to recover the perturbative integrability  of these models entirely from the underlying Lie algebra in a universal way; in this light, this work represents a further step in that direction. We expect that the answers, at least for the class of ADE affine Toda models, should be hidden in the root system underlying the theory \cite{Dorey:1990xa}, from which the on-shell momenta at the pole positions are projected.

\vskip 10pt
\noindent
{\bf Acknowledgments}\\[3pt]
%
% \noindent
We thank Ben Hoare for useful discussions. This work has received funding from the  European Union's  Horizon  2020  research  and  innovation programme under the Marie  Sk\l odowska-Curie  grant  agreement  No.~764850 \textit{``SAGEX''}, and from the STFC under consolidated grant ST/T000708/1 “Particles, Fields and Spacetime”.

\vskip 10pt

\appendix

\section{Even order singularities in the bootstrapped S-matrix}
\label{Appendix_S_matrix_expansion}

The S-matrices of simply-laced affine Toda theories can be written in terms of building blocks \cite{Braden:1989bu}
\begin{equation}
\label{brink_definition_curly_bracket}
\{ y \} = \frac{(y+1) (y-1)}{(y-1+B) (y+1-B)} 
\end{equation}
where
$$
(y)\equiv \frac{\sinh \bigl( \frac{\theta}{2} + \frac{i \pi y}{2h} \bigr)}{\sinh \bigl( \frac{\theta}{2} - \frac{i \pi y}{2h} \bigr)} \hspace{6mm} \text{and} \hspace{6mm} B \equiv \frac{1}{2 \pi} \frac{\beta^2}{1+\frac{\beta^2}{4 \pi}}.
$$
In this expression 
$\theta$  is the difference between the rapidities of the interacting particles and $\beta$ is the Lagrangian coupling present in~\eqref{Toda_theory_lagrangian_defined_in_terms_of_roots}. The S-matrix is a product over such building blocks (also called bricks) which can be written schematically as
$$
S_{ab}(\theta)= \prod_{y=1}^{h-1} \{ y \}^{N_{ab}^{(y)}}.
$$
The exponents $N_{ab}^{(y)}$ of the different bricks are non-negative integer coefficients depending on the point $y$ at which they are evaluated, and also on the particles $a$ and $b$ scattered; they can be expressed in terms of root system data \cite{Dorey:1990xa}.
A feature of simply-laced affine Toda theories is that, for a given S-matrix element $S_{ab}$, the non-zero multiplicities of its building blocks are all located at even or odd values of $y$. In other words, one of the two following conditions has to hold
 \begin{itemize}
    \item $N_{ab}^{(2k)}=0 \ \forall \ k \in \mathbb{N}$;
    \item $N_{ab}^{(2k+1)}=0 \ \forall \ k \in \mathbb{N}$. 
\end{itemize}   
This fact, already noted in~\cite{Braden:1989bu}, was explained in~\cite{Dorey:1990xa,Dorey:1991zp} and relies on the properties of root systems.  
Each building block $\{ y \}$ has two simple poles, at purely imaginary
rapidities $\theta=\frac{i \pi}{h} (y \pm 1)$. In this paper we are interested in studying even-order
poles due to bricks of equal multiplicity that touch each other, i.e.\  to the situation $N_{ab}^{(x-1)}=N_{ab}^{(x+1)}=N$. This is in fact the only way that even-order poles can arise \cite{Dorey:1991zp}.
In particular we focus on singular terms arising from the S-matrix expansion around such pole positions up to order $\beta^4$ in the coupling expansion.

The  singular expansion of the bricks
around the pole position $\theta_0=\frac{\pi x}{h}$ 
up to order $B^2$ (which includes all  powers of $\beta$ less than or equal to $\beta^4$) has the following form
\begin{equation}
\label{brink_and_remaining_S_matrix_expansion}
\begin{split}
&\{x-1\}=1+B \Bigl(\frac{a_-}{\theta-i\theta_0}+b_-+\dots\Bigr)+B^2 \left(\frac{c_-}{\theta-i\theta_0}+\dots\right)+\dots,\\
&\{x+1\}=1+B \Bigl(\frac{a_+}{\theta-i\theta_0}+b_++\dots \Bigr)+B^2 \left(\frac{c_+}{\theta-i\theta_0}+\dots\right)+\dots,\\
&\prod_{y \ne x \pm1} \{ y \}\bigl|_{\theta=i\theta_0}^{N_{ab}(y)}=1+e B+\dots
\end{split}
\end{equation}
where the letters $a_-$, $a_+$, $b_-$, $b_+$ and $e$ label
the coefficients of the double expansion, and we have omitted all terms which cannot contribute to the singular part of the S-matrix at $\theta=i\theta_0$, up to order $B^2$, in any product of these blocks.
By expanding the  blocks first with respect to $B$ and then around the pole $\theta=i\theta_0$, it is easy to check the following identities:
\begin{equation}
\label{identities_on_the_terms_of_the_brink_expansion}
\begin{split}
a_-=-a_+&= \frac{i \pi}{h},\\
(c_- + c_+)  &= \frac{i \pi}{h}(b_- - b_+).
\end{split}
\end{equation}
The only way in which the term $eB$ in \eqref{brink_and_remaining_S_matrix_expansion} can contribute to the pole at the power of $B$ that we are interested in is when it multiplies the quantities proportional to $\frac{a_-}{\theta-i\theta_0}$ and $\frac{a_+}{\theta-i\theta_0}$ 
$$
\{x-1\}^N \{x+1\}^N \prod_{\substack{y \ne x \pm1 \\ \text{step $2$}}} \{ y \}^{N_{ab}(y)} \leadsto N \frac{eB^2}{\theta-i\theta_0} (a_+ + a_-),
$$
and this is zero by the first identity in \eqref{identities_on_the_terms_of_the_brink_expansion}.
Taking into account that $a_-=-a_+=\frac{i \pi}{h}$ the singular part of the S-matrix close to the $2N$-order pole, up to order $B^2$, is then given by
\begin{equation}
\begin{split}
&S_{ab}(\theta)\sim \{x-1\}^N \ \{x+1\}^N\\
&=B^2 \biggl[ \frac{\pi^2}{h^2}\frac{N}{(\theta-i\theta_0)^2} \ + \ \frac{N}{\theta-i\theta_0} \Bigl(  (c_- + c_+)  - \frac{i \pi}{h} (b_- - b_+ )\Bigr)  \biggr] + \ldots
\end{split}
\end{equation}
We see that at  order $B$ no singular term appears. Moreover the coefficient  of $\frac{1}{\theta-i\theta_0}$ at order $B^2$ vanishes by the second equality in \eqref{identities_on_the_terms_of_the_brink_expansion}.
Expanding $B$ in terms of the Lagrangian coupling $\beta$
$$
B= \frac{\beta^2}{2 \pi} + O(\beta^4)
$$
we conclude that up to order $\beta^4$, the only singular term is a second-order pole, with
all simple pole contributions cancelling due to the structure of the coefficients of the building blocks:
\begin{equation}
\label{S_matrix_expansion_obtained_using_bootstrap_on_the_pole}
S_{ab}(\theta) =  \Bigl(\frac{\beta^2}{2h} \Bigr)^2  \frac{N}{(\theta-i\theta_0)^2} + \ldots
\end{equation}
Comparing~\eqref{S_matrix_expansion_obtained_using_bootstrap_on_the_pole} with the Laurent expansion~\eqref{Expansion_of_S_around_an_arbitrary_2N_order_pole} confirms that the values of the coefficients extracted from the bootstrapped S-matrix are $a_1=b_1=0$ and $a_2=N$. 
Higher-order singularities of the form $(\theta-i\theta_0)^{-p}$, with $2<p \le 2N$, are also present in the full Laurent expansion of the S-matrix for $N>1$,
but they contain powers of the coupling greater than $4$ and we exclude them from the present analysis since we are only considering contributions generated by one-loop diagrams. 

\section{Deformations of integration surfaces}
\label{Appendix_contour_integrals}

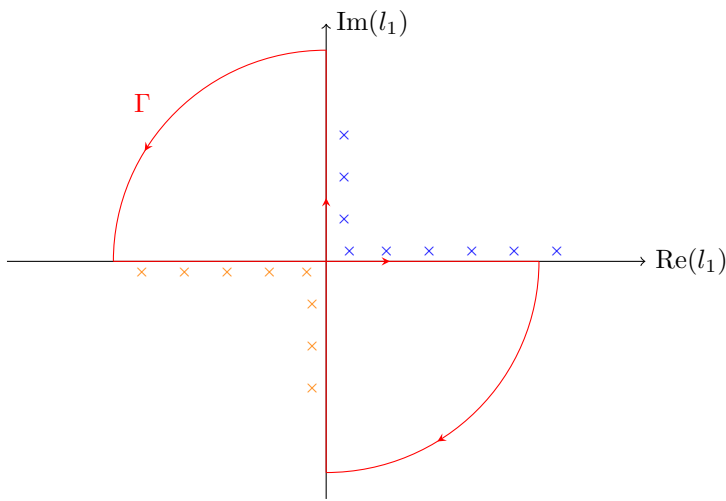
\begin{figure}
\medskip
\begin{center}
\begin{tikzpicture}
\tikzmath{\y = 0.7;}

\draw[->][] (-6*\y,0*\y) -- (6*\y,0*\y);
\draw[->][] (0*\y,-4.5*\y) -- (0*\y,4.5*\y);

\draw[directed][red] (4*\y,0*\y) arc(0:-90:4*\y);
\draw[directed][red] (0*\y,-4*\y) -- (0*\y,4*\y);
\draw[directed][red] (0*\y,4*\y) arc(90:180:4*\y);
\draw[directed][red] (-4*\y,0*\y) -- (4*\y,0*\y);
\filldraw[red] (-3.8*\y,3*\y)  node[anchor=west] {\small{$\Gamma$}};

\filldraw[] (0*\y,4.5*\y)  node[anchor=west] {\small{$\Im(l_1)$}};
\filldraw[] (6*\y,0*\y)  node[anchor=west] {\small{$\Re(l_1)$}};

\filldraw[blue] (0*\y,2.4*\y)  node[anchor=west] {\tiny{$\times$}};
\filldraw[blue] (0*\y,1.6*\y)  node[anchor=west] {\tiny{$\times$}};
\filldraw[blue] (0*\y,0.8*\y)  node[anchor=west] {\tiny{$\times$}};
\filldraw[blue] (0.1*\y,0.2*\y)  node[anchor=west] {\tiny{$\times$}};
\filldraw[blue] (0.8*\y,0.2*\y)  node[anchor=west] {\tiny{$\times$}};
\filldraw[blue] (1.6*\y,0.2*\y)  node[anchor=west] {\tiny{$\times$}};
\filldraw[blue] (2.4*\y,0.2*\y)  node[anchor=west] {\tiny{$\times$}};
\filldraw[blue] (3.2*\y,0.2*\y)  node[anchor=west] {\tiny{$\times$}};
\filldraw[blue] (4*\y,0.2*\y)  node[anchor=west] {\tiny{$\times$}};

\filldraw[orange] (-0.6*\y,-2.4*\y)  node[anchor=west] {\tiny{$\times$}};
\filldraw[orange] (-0.6*\y,-1.6*\y)  node[anchor=west] {\tiny{$\times$}};
\filldraw[orange] (-0.6*\y,-0.8*\y)  node[anchor=west] {\tiny{$\times$}};
\filldraw[orange] (-0.7*\y,-0.2*\y)  node[anchor=west] {\tiny{$\times$}};
\filldraw[orange] (-1.4*\y,-0.2*\y)  node[anchor=west] {\tiny{$\times$}};
\filldraw[orange] (-2.2*\y,-0.2*\y)  node[anchor=west] {\tiny{$\times$}};
\filldraw[orange] (-3*\y,-0.2*\y)  node[anchor=west] {\tiny{$\times$}};

\filldraw[orange] (-3.8*\y,-0.2*\y)  node[anchor=west] {\tiny{$\times$}};

\end{tikzpicture}
\end{center}

\caption{Solutions $l_1^{(+)}$ (in blue) and $l_1^{(-)}$ (in orange) in the case $u=0$ for different real values of $l_0$. The red contour $\Gamma$ does not enclose poles of the propagator.}
\label{Poles_of_propagator_case_u_zero_appendix}
\end{figure}

In this appendix, we justify why the loop integrations in section~\ref{Singularities_from_Feynman_diagrams_main_section} must be performed over regions with purely imaginary $l_1$. Let us consider a generic loop propagator with a denominator
\begin{equation}
\label{denominator_f_of_propagator_appendix}
    f(l)\equiv(P+l)^2-m^2_A+i \epsilon,
\end{equation}
where $P$ is a vector determined by the external kinematics, $m_A$ is the mass of the propagating particle and $l=(l_0,l_1)$ is the loop momentum over which we need to integrate. For purely imaginary values of the rapidities of the external particles, $P$ will be given by
\begin{equation}
    P=m_P(\cos{u}, i\sin{u}),
\end{equation}
where $m_P$ is a positive quantity, corresponding to the length of the vector $P$, and depends on the external momenta. The parameter $u$ is instead a real number belonging to the interval $[0, 2\pi)$. To find the zeros of $f(l)$, corresponding to the poles of the propagator,
we need to solve the equation $f(l)=0$; we choose to solve this equation, which can be written as 
\begin{equation}
\label{equation_fl_equal_zero_appendix}
    l_1^2 +2 i m_P \sin{u} \ l_1 +(m_A^2-m_P^2-2m_P \cos{u} \ l_0 - l_0^2 - i \epsilon)=0,
\end{equation}
in the variable $l_1$. 
From~\eqref{equation_fl_equal_zero_appendix} we see that, for $l_0$ real, there are no purely imaginary solutions $l_1$. This is evident by the fact that, if $l_0 \in \mathbb{R}$ and $i l_1 \in \mathbb{R}$, the only imaginary term in~\eqref{equation_fl_equal_zero_appendix} is $i \epsilon$. On the contrary, for $l_0 \in \mathbb{R}$, there can be in general solutions for $l_1$ on the real spatial axis. We show this in some more detail. The two solutions of~\eqref{equation_fl_equal_zero_appendix} are given by
\begin{equation}
    l_1^{(\pm)}=-i m_P \sin{u} \pm \sqrt{(m_P \cos{u}+l_0)^2 - m_A^2 + i\epsilon},
\end{equation}
which expanded in small $\epsilon$ become
\begin{equation}
\label{solutions_lplusminus_u_nonzero_appendix}
    l_1^{(\pm)}=-i m_P \sin{u} \pm \sqrt{(m_P \cos{u}+l_0)^2 - m_A^2} \Bigl(1+\frac{i\epsilon}{(m_P \cos{u}+l_0)^2 - m_A^2} \Bigr).
\end{equation}
Let us consider the case $u=0$ first; this should be imagined as the case in which the external particles entering the loop have zero rapidities. In this case, the solutions are
\begin{equation}
\label{solutions_for_l0_l1_in_the_case_u_equal_zero_appendix}
    l_1^{(\pm)}= \pm \sqrt{(m_P+l_0)^2 - m_A^2} \cdot \Bigl(1+\frac{i\epsilon}{(m_P+l_0)^2 - m_A^2} \Bigr)
\end{equation}
and are depicted in figure~\ref{Poles_of_propagator_case_u_zero_appendix} for different real values of $l_0$. The markers close to the real axis correspond to values of $l_0$ at which $(m_P+l_0)^2 - m_A^2$ is positive, while the markers close to the imaginary axis correspond to the case in which such a quantity is negative.
The contour $\Gamma$ in figure~\ref{Poles_of_propagator_case_u_zero_appendix} does not enclose zeros of~\eqref{denominator_f_of_propagator_appendix}; therefore, if we integrate $1/f(l)$ over $\Gamma$, we obtain zero. If we think to have a function
$$
F(l)=\frac{1}{f_1(l) f_2(l) \ldots},
$$
where each $f_j(l)$ is the denominator of a certain propagator, then at zero external rapidities (where all the different angles $u$, one for each propagator, are zero) it holds that
$$
\oint_\Gamma F(l)=0.
$$
If we close the contour $\Gamma$ at infinity so that the two arc contributions are suppressed then it has to hold 
$$
\int_{-\infty}^{+\infty} dl_1 F(l) = \int_{+i \infty}^{-i \infty} dl_1 F(l).
$$
This implies that the Feynman diagram having as integrand $F(l)$ can be equivalently be written as
\begin{equation}
\label{equivalent_integration_surfaces_at_u_equal_to_zero_appendix}
    D=\int_{-\infty}^{+\infty} d l_0 \int_{-\infty}^{+\infty} dl_1 F(l)= \int_{-\infty}^{+\infty} d l_0 \int_{+i \infty}^{-i \infty} dl_1 F(l).
\end{equation}

If we continue from our starting-point to imaginary values of the rapidities of the external particles then, at real values of $l_0$, poles can cross the real axis of $l_1$. For example, if we consider the solution $l_1^{(+)}$ in~\eqref{solutions_lplusminus_u_nonzero_appendix} for $u$ slightly bigger than $0$ and in the case $(m_P \cos{u}+l_0)^2-m_A^2>0$, then we note that $\Im(l_1^{(+)})<0$. 
When $u$ is turned on, the blue markers close to the real axis in figure~\ref{Poles_of_propagator_case_u_zero_appendix} cross the real axis and acquire a negative imaginary part. As a consequence of this fact, if we want to analytically continue the first equality in~\eqref{equivalent_integration_surfaces_at_u_equal_to_zero_appendix} to $u\ne 0$, integrating over real $l_0$, the integration path of $l_1$ cannot be real anymore. This path has to be deformed below the positive side of the real axis in such a way to avoid the poles. 
Due to this fact, the first equality in~\eqref{equivalent_integration_surfaces_at_u_equal_to_zero_appendix} is not true anymore. However, since the poles never cross the imaginary axis, the second equality in~\eqref{equivalent_integration_surfaces_at_u_equal_to_zero_appendix} still holds and the Feynman integral can still be consistently reproduced by integrating over purely imaginary $l_1$.

%\newpage

\end{document}